\documentclass[iop]{emulateapj}
\usepackage{epsfig}
\usepackage{amsmath}
\usepackage{array,multirow}
\numberwithin{equation}{subsection}
\usepackage{amssymb}
\usepackage{threeparttable}
\usepackage{amsmath}

\usepackage{longtable}
\LongTables
\usepackage{color}
\usepackage{hyperref}
\usepackage{bm,url}
\usepackage{hyperref}
\hypersetup{
    colorlinks=true,
    linkcolor=blue,
    filecolor=magenta,      
    urlcolor=cyan,
}

\providecommand{\N[1]}{\ensuremath{N_{\rm #1}}}
\providecommand{\nh}{\ensuremath{\N{H}}}
\providecommand{\nhi}{\ensuremath{\N{H\,I}}}
\providecommand{\meh}{\ensuremath{[{\rm M/H}]}} 
\providecommand{\logu}{\ensuremath{\log U}}
\providecommand{\logulim}{\ensuremath{\log U_{\rm Type1}}}
\providecommand{\aratio}{\ensuremath{[{\rm \alpha/Al}]}}
\providecommand{\zlls}{\ensuremath{z_{\rm LLS}}}
\providecommand{\zqso}{\ensuremath{z_{\rm QSO}}}
\providecommand{\cm[1]}{\ensuremath{{\rm \,cm}^{#1}}}

\providecommand{\kms}{\ensuremath{\,{\rm km\,s}^{-1}}}

\providecommand{\xh[1]}{\ensuremath{[\mathrm{#1}/\mathrm{H}]}}

\usepackage{array}
\usepackage{booktabs}
\usepackage{multirow}
\newcommand{\head}[1]{\textnormal{\textbf{#1}}}

\begin{document}
\title{Predominantly Low Metallicities Measured\\ in a Stratified Sample of Lyman Limit Systems at $z=3.7$} 
\author{Ana Glidden,\altaffilmark{1} Thomas J. Cooper,\altaffilmark{1} Kathy L. Cooksey,\altaffilmark{2} Robert A. Simcoe,\altaffilmark{1,3}  and John M. O'Meara\altaffilmark{4}} 

\altaffiltext{1}{Massachusetts Institute of Technology, 77 Massachusetts Ave, Cambridge, MA 02139, USA; \texttt{aglidden@mit.edu,tjcooper@mit.edu,simcoe@space.mit.edu}}

\altaffiltext{2}{Department of Physics \& Astronomy, University of Hawai`i at Hilo, 200 West K\=awili Street, Hilo, HI 96720, USA; \texttt{kcooksey@hawaii.edu}}

\altaffiltext{3}{MIT-Kavli Center for Astrophysics and Space Research, Cambridge, MA 02139, USA}

\altaffiltext{4}{Department of  Physics, Saint Michael's College, One Winooski Park, Colchester, VT 05439, USA; \texttt{jomeara@smcvt.edu}}

\shorttitle{Low-Metallicity Lyman Limit Systems II}\shortauthors{Glidden et al.}

\begin{abstract}

We measured metallicities for 33 $z=3.4$--4.2 absorption line systems drawn from a sample of \ion{H}{1}-selected-Lyman limit systems (LLSs) identified in Sloan Digital Sky Survey (SDSS) quasar spectra and stratified based on metal line features. We obtained higher-resolution spectra with the Keck Echellette Spectrograph and Imager, selecting targets according to our stratification scheme in an effort to fully sample the LLS population metallicity distribution. We established a plausible range of \ion{H}{1} column densities and measured column densities (or limits) for ions of carbon, silicon, and aluminum, finding ionization-corrected metallicities or upper limits. Interestingly, our ionization models were better constrained with enhanced $\alpha$-to-aluminum abundances, with a median abundance ratio of $\aratio=0.3$. Measured metallicities were generally low, ranging from $\meh=-3$ to $-1.68$, with even lower metallicities likely for some systems with upper limits. Using survival statistics to incorporate limits, we constructed the cumulative distribution function (CDF) for LLS metallicities. Recent models of galaxy evolution propose that galaxies replenish their gas from the low-metallicity intergalactic medium (IGM) via high-density \ion{H}{1} ``flows" and eject enriched interstellar gas via outflows. Thus, there has been some expectation that LLSs at the peak of cosmic star formation ($z\approx3$) might have a bimodal metallicity distribution. We modeled our CDF as a mix of two Gaussian distributions, one reflecting the metallicity of the IGM and the other representative of the interstellar medium of star-forming galaxies. This bimodal distribution yielded a poor fit. A single Gaussian distribution better represented the sample with a low mean metallicity of $\meh\approx -2.5$.

\end{abstract}

\section{INTRODUCTION}

It has been hypothesized that the large star formation rates seen in galaxies at redshift $z\approx 2$--3 reflect direct gas accretion from the intergalactic medium (IGM) onto galactic disks.  According to this picture, gas transport occurs along cold filaments ($T\lesssim10^5\,$K) that do not shock at the halo's virial radius. The filaments provide the fuel supply for star formation and can be replenished on dynamical timescales \citep{Dekela, Dekelb, Nelson}.

In numerical simulations, such ``cold flows'' are not independently luminous, but are optically thick in \ion{H}{1} and so may easily be seen in absorption \citep{Fumagalli11b,vandeVoort}.  Observationally, they have properties similar to the Lyman limit systems (LLSs) often seen in QSO spectra \citep[e.g.,][]{Sargent}. LLSs are canonically defined by their column density ($\nhi \ge 10^{17.5}\cm{-2}$), which is large enough to absorb quasar light ($\tau_{912}\geq1$) blueward of the Lyman limit at 912\,\AA\ (1\,Ryd), redshifted to the absorber frame. Yet, because galaxies accrete from the diffuse, low-metallicity (if not primordial) IGM, we expect that the LLSs representing cold flows should have low heavy-element abundance.

Empirically, the majority of LLSs (which are selected in \ion{H}{1}) also exhibit heavy-element absorption lines and were traditionally not considered likely candidates for low-metallicity gas.  However, the
existence of heavy-element absorption does not constitute {\em prima facie} evidence of a high abundance.  Instead, detailed modeling is required to determine the heavy-element content, because the LLSs are optically thick yet substantially ionized \citep[e.g.,][]{Prochaska1999}, and in most cases, even their \ion{H}{1} column density is very poorly constrained by the absorption data.  Nevertheless, the potential connection between low-metallicity LLSs and the predicted cold flows has motivated us and other groups to study LLS abundances at both low and high redshifts.

At $z<1$, \citet{Lehner} uncovered evidence for a bimodal distribution of LLS metallicities using the Cosmic Origins Spectrograph on the {\it Hubble Space Telescope}. Although the absorbers chosen in their sample range from $16.2<\log\nhi<18.5$, and the distribution is tilted toward the low end of that column density range such that many sub-LLSs are included, their measurements and simulations have withstood repeated observation (Lehner 2016, private communication).  According to their favored interpretation, the high-metallicity branch of the LLS
distribution ($[{\rm M/H}]\approx-0.3$)\footnote{The square-bracket notation for metallicity and abundances, e.g., [X/H], is relative to solar. Thus, for some element X: $[{\rm X/H}] = \log(\N{X}/\N{H})-\log(\N{\rm X,\odot}/\N{\rm H,\odot})$. We use the notion \meh, to indicate total metallicity assuming all elements have the same relative abundance, i.e., $\meh = [{\rm X/H}]$ for all X.} is associated with feedback from nearby galaxies, while the low end ($[{\rm M/H}]\approx-1.6$) is associated with accretion from the IGM. 

In support of this interpretation, \citet{Kacprzak2012a} associated a $z=0.7$, low-metallicity absorber with a nearby, solar-metallicity galaxy and concluded that the gas detected in absorption is likely accreting. \citet{Kacprzak2012b} found that the azimuthal-angle distribution (measured from the galaxy major axis) of \ion{Mg}{2} $\lambda\lambda 2796,2803$ absorption around galaxies also contains a bimodality, largely driven by star-forming galaxies, suggesting that the bimodality reflects gas inflow and outflow.  \citet{2011ApJ...743...10B} similarly find strong azimuthal dependence in the equivalent widths of \ion{Mg}{2} absorbers around blue disk galaxies with strong absorbers preferentially found along galaxy minor axes at small impact parameters. \citet{2014ApJ...784..108B} compare models to the observed \ion{Mg}{2} distribution and attribute the azimuthal dependence to outflowing winds. 

However, other observations suggest a more complex picture. For example, \citet{2012ApJ...747L..26R} find six instances of cool, metal-rich gas accreting onto $z\sim0.5$ galaxies, possibly from recycling gas or dwarf satellites, reinforcing the notion that not all inflows with the potential to trigger star formation are cold flows from the IGM \citep{2010MNRAS.406.2325O,2012ARA&A..50..491P}. More recent results at $z<1$ also indicate that the degree of metallicity bimodality reported in \citep{Lehner} depends upon the range of $\nhi$ considered \citep{2016arXiv160802584W}. While sub-LLSs show a bimodality, above $\nhi=17.2$, the high-metallicity branch is suppressed. This is consistent with the recent work of \citet{2016arXiv160802588L}, who found low metallicities for sub-LLSs and LLSs alike at $z>2$.

Still, simulations by \citet{Neistein} predict that cold-flow accretion should increase with redshift as $\dot{M} \propto (1+z)^{2.5}$ and hence be much more prevalent at $z>2$ than in the local universe.  Select analyses of individual LLSs at higher redshift suggested that some indeed have very low abundances (or components with low abundances), consistent with being randomly drawn from the IGM \citep{Levshakov2003,Fumagalli,Crighton2013,Crighton2016}.

With this in mind, in \defcitealias{Cooper}{Paper I}\citet[][hereafter Paper I]{Cooper} we analyzed a well-defined statistical sample of 17 LLSs at $z=3.2$--4.4, directly overlapping with the $z=2$--4 epoch where cold flows should be most common. The sample was uniformly selected based on an \ion{H}{1} optical depth of $\tau_{912}\geq2$ (equivalent to $\nhi \ge 10^{17.5}\cm{-2}$) at the Lyman limit, using a large survey of QSO spectra \citep{Prochaska}. This is not strictly identical to the definition of LLSs ($\tau_{912}\geq1$), but the \citeauthor{Prochaska} catalog of 194 systems is complete to $\log \nhi > 17.3$ and includes both LLSs and damped Ly$\alpha$ systems (DLAs). Our additional selection criterion excluded sightlines exhibiting metal absorption in their SDSS spectra, eliminating about 50\% of high-redshift LLSs observed with SDSS.  In \citetalias{Cooper}, we found metallicities in this ``metal-poor'' sample ranging from $\meh=-2$ to $\meh<-3$ in the subsample. Factoring in the subsample selection, it was extrapolated that 28--40\% of the SDSS LLS population at $z\approx3.7$ has metallicity consistent with the IGM and hence potentially represents cold-flow accretion. In \citetalias{Cooper}, we also analyzed ten LLSs (``metal-blind sample'') at $z\sim3.0$ from the blind LLS survey of \citet{2013ApJ...775...78F}.

\citet{Fumagalli15} also recently examined a sample of 157 LLSs at $z=1.8$--4.4, drawn from a combination of spectra observed for other programs and archival data in the public domain.  Like \citetalias{Cooper}, these authors found predominantly low metallicities for the LLS population; because of their larger sample size and no explicit bias toward lower-metallicity absorbers, they were also able to rule out a bimodal distribution similar to that at low redshift. (There exist a small number of metal-rich absorbers in their sample, but these are mostly at higher, sub-DLA \ion{H}{1} column densities.)
 
Here we analyze a sample of 33 high-redshift LLSs along SDSS quasar sightlines, using the procedural framework developed in \citetalias{Cooper}. Unlike the metal-poor sample in \citetalias{Cooper} these sightlines were {\it not} subject to exclusion on the basis of detected absorption lines, so they form a pure \ion{H}{1}\!-selected sample from a large, well-defined survey (SDSS) and are highly representative of the LLS population as a whole. 

In Section \ref{sec.obs}, we detail our observations. With ionization modeling and a Markov-Chain Monte-Carlo (MCMC) analysis described in Section \ref{sec.analysis}, we determine the metallicity of the absorbing gas for each system. Low metallicity implies that the gas is a viable candidate for cold accretion, while high metallicity implies that the gas has been polluted with heavy elements produced in stars from a presumed nearby galaxy. In Section \ref{sec.results}, we compare measured metallicities with ionization properties and \ion{H}{1} column densities and consider evidence that measured aluminum abundances are not consistent with other elements, perhaps due to different nucleosynthetic origins. Finally, in Section \ref{sec:pdf}, we create an LLS metallicity distribution to determine what fraction of our LLSs trace gas directly drawn (probably) from the IGM and if an abundance bimodality exists.

Throughout, we adopt a standard cosmology: $\Omega_m = 0.28$, $\Omega_\Lambda = 0.72$, $H_0 = 70\, {\rm km}\, {\rm s}^{-1}\, {\rm Mpc}^{-1}$ \citep{Hinshaw}.

\begin{table*}
\centering
\begin{threeparttable}
\caption{Details for the Keck/ESI Observations \label{Observations}}
\begin{tabular}{ccccccccccc}
\toprule[0.75pt]
\toprule[0.25pt]
\multicolumn{1}{c}{\head{QSO}} &  \multicolumn{1}{c}{\head{ R.A.}} &  \multicolumn{1}{c}{\head{Dec.}} & 
\multicolumn{1}{c}{\head{ $\bm{\zqso}$ }} &\multicolumn{1}{c}{\head{ $\bm{\zlls}$ }} &  \head{Tier} & \multicolumn{1}{c}{\head{ Exp (s) } }&  \multicolumn{1}{c}{\head{ $\bm{\log\nhi}$ }} & 
\multicolumn{1}{c}{\head{$\bm{\meh}$}} &  \multicolumn{1}{c}{\head{$\bm{\logu}$}}&  \multicolumn{1}{c}{\head{$\bm{\aratio}$}}\\[5pt]
\hline\\
J011351--093551 &  01:13:51.96  &  --09:35:51.0 &  3.668 &  3.617  & 3 &  2$\times$900           &17.80--19.10& --2.07$_{-0.11}^{+0.13}$ &--2.07$_{-0.09}^{+0.08}$ & 0.08$\pm0.11$\\[5pt] 
J034402--065300 &  03:44:02.85  & --06:53:00.6  &  3.957 &  3.843  & 1 &  2$\times$975           &17.80--19.40& --3.00$_{-0.19}^{+0.26}$ &--2.07$_{-0.13}^{+0.12}$& $>$0.03\\[5pt] 
J075103+424211  &  07:51:03.95  & +42:42:11.6   &  4.163 &  4.051  & 1 &  1500                         &17.80--18.60& --2.50$_{-0.13}^{+0.15}$& --2.40$\pm0.08$& $>$0.48\\[5pt]
J081039+345730  &  08:10:39.79  & +34:57:30.9   &  3.772 &  3.506  & 3 &  2$\times$1150         &17.80--19.05& --1.96$_{-0.18}^{+0.13}$& --2.00$_{-0.09}^{+0.10}$& 0.17$_{-0.14}^{+0.13}$\\[5pt]      
J081809+321912  &  08:18:09.56  & +32:19:12.8   &  3.785 &  3.655  & 2 &  2$\times$1750         &17.80--19.30& --2.25$_{-0.17}^{+0.18}$& --2.24$\pm0.10$& $>$0.42\\[5pt]    
J081855+095848  &  08:18:55.78  & +09:58:48.0   &  3.674 &  3.531  & 3 &  600, 900                   &17.80--18.60& --2.33$_{-0.14}^{+0.16}$& --1.41$\pm0.24$& 0.08$\pm0.13$\\[5pt]    
J082340+342753  &  08:23:40.48  & +34:27:53.0   &  4.248 &  4.190  & 2 &  2$\times$1350         &17.80--19.25& $<$--2.75& $>$--2.28 & $>$--0.83\\[5pt]  
J083941+031817  &  08:39:41.45  & +03:18:17.0   &  4.248 &  4.154  & 3 &  1500, 1250               &17.80--18.55& --2.35$_{-0.17}^{+0.21}$& --2.46$_{-0.10}^{+0.09}$&$>$0.33\\[5pt]
J100412+292121  &  10:04:12.42  & +29:21:21.5   &  3.694 &  3.566  & 1 &  2$\times$1150         &17.80--19.00& --2.96$\pm0.47$& --2.04$_{-0.27}^{+0.29}$& \nodata\\[5pt]
J101347+065015  &  10:13:47.29  & +06:50:15.6   &  3.792 &  3.490  & 2 &  2$\times$1250         &17.80--19.40& --2.08$_{-0.25}^{+0.20}$& --2.33$_{-0.11}^{+0.13}$ & 0.29$_{-0.15}^{+0.14}$\\[5pt]   
J103018+164633  &  10:30:18.43  & +16:46:33.0   &  3.988 &  3.802  & 3 &  2$\times$800,1200    &17.80--19.90& --2.25$_{-0.17}^{+0.21}$& --2.13$_{-0.12}^{+0.11}$& 0.71$\pm0.18$\\[5pt] 
J103048+391234  &  10:30:48.24  & +39:12:34.3   &  3.735 &  3.482  & 2 &  2$\times$1500         &17.80--19.35& $<$--2.29&$>$--3.00& \nodata\\[5pt] 
J104057+514505  &  10:40:57.68  & +51:45:05.8   &  4.047 &  3.931  & 1 &  2$\times$975           &17.80--19.45& $<$--2.61& $>$--3.00 &\nodata\\[5pt]    
J105830+333859  &  10:58:30.03  & +33:38:59.3   &  3.833 &  3.641  & 2 &  2$\times$1350         &17.80--18.55& $<$--2.46& $>$--3.00& \nodata\\[5pt]        
J110236+460101  &  11:02:36.79  & +46:01:01.3   &  3.845 &  3.595  & 1 &  1250, 2000               &17.80--19.15& $<$--2.58& $>$--3.00&\nodata\\[5pt]
J111957+281354  &  11:19:57.10  & +28:13:54.1   &  4.100 &  3.691  & 3 &  2$\times$1150         &17.80--19.10& $<$--2.61& $>$--3.00&\nodata\\[5pt] 
J113608+250322  &  11:36:08.53  & +25:03:22.1   &  3.625 &  3.559  & 1 &  2$\times$1350         &17.80--19.65&--2.33$_{-0.19}^{+0.24}$& --2.37$_{-0.12}^{+0.11}$&$>$--0.57\\[5pt] 
J114713+362702  &  11:47:13.01  & +36:27:02.4   &  3.794 &  3.393  & 1 &  2$\times$1250         &17.80--19.95&$<$--2.15& $>$--3.00& \nodata\\[5pt]     
J121058+182119\tablenotemark{a}  &  12:10:58.56  & +18:21:19.1   &  3.881 &  3.732  & 2 &  2$\times$1350         &17.40--17.65& --3.54$_{-0.24}^{+0.31}$& --1.41$_{-0.21}^{+0.17}$&$>$--1.36\\[5pt]    
J122000+254230  &  12:20:00.83  & +25:42:30.7   &  4.034 &  3.921  & 3 &  2$\times$550           &17.80--18.90&--2.76$_{-0.16}^{+0.17}$& --2.17$\pm0.12$& $>$--0.77\\[5pt] 
J122027+261903  &  12:20:27.96  & +26:19:03.5   &  3.697 &  3.508  & 1 &  2$\times$550           &17.80--18.50& $<$--2.51& $>$--3.00& \nodata\\[5pt] 
J131453+080456  &  13:14:53.03  & +08:04:56.6   &  3.733 &  3.509  & 3 &  2$\times$1150         &19.75--19.90& --1.68$\pm0.08$& --2.38$\pm0.06$& 0.13$\pm 0.09$\\[5pt]    
J140248+014634  &  14:02:48.07  & +01:46:34.1   &  4.161 &  3.796  & 2 &  1150, 1500               &17.80--18.85& --2.86$_{-0.16}^{+0.18}$& --1.83$_{-0.10}^{+0.11}$& $>$0.31\\[5pt]
J141831+444937  &  14:18:31.70  & +44:49:37.5   &  4.312 &  4.122  & 1 &  2$\times$1250         &17.80--19.50&--2.48$_{-0.14}^{+0.17}$& --2.40$_{-0.09}^{+0.08}$& 0.24$\pm0.14$\\[5pt]    
J144144+472003\tablenotemark{a}  &  14:41:44.76  & +47:20:03.2   &  3.633 &  3.593  & 1 &  2$\times$900           &17.20--17.60&--3.12$_{-0.23}^{+0.25}$& --1.32$_{-0.16}^{+0.15}$& $>$--1.18\\[5pt] 
J144213+391856  &  14:42:13.09  & +39:18:56.0   &  3.627 &  3.558  & 3 &  550, 850                   &17.80--19.40& $<$--2.15& $>$--3.00& \nodata\\[5pt]   
J144335+334859  &  14:43:35.16  & +33:48:59.8   &  3.657 &  3.419  & 3 &  2$\times$975           &19.60--19.70& --1.92$\pm0.08$& --2.34$\pm0.06$& 0.07$\pm0.09$\\[5pt]
J144542+490248\tablenotemark{b}  &  14:45:42.76  & +49:02:48.9   &  3.875 &  3.660  & 2 &  1$\times$750 &17.80--18.85& --2.29$_{-0.14}^{+0.17}$& --2.38$_{-0.09}^{+0.08}$& $>$--0.37\\[5pt]  
J145243+015430  &  14:52:43.61  & +01:54:30.7   &  3.908 &  3.749  & 2 &  2$\times$1250         &17.80--18.75& --2.46$\pm0.15$& --2.37$_{-0.09}^{+0.08}$& $>$0.19\\[5pt]    
J151352+204057  &  15:13:52.09  & +20:40:57.6   &  3.717 &  3.452  & 2 &  2$\times$1350         &17.80--18.80& --2.48$_{-0.15}^{+0.16}$& --2.12$_{-0.09}^{+0.10}$&$>$0.37\\[5pt] 
J152436+212309  &  15:24:36.08  & +21:23:09.1   &  3.607 &  3.464  & 3 &  2$\times$300           &17.80--19.00&  $<$--2.52& $>$--2.82& $>$--0.79\\[5pt]  
J152652+405126  &  15:26:52.76  & +40:51:26.6   &  3.713 &  3.660  & 3 &  2$\times$1150         &17.80--18.90& --2.71$_{-0.38}^{+0.41}$& --1.72$_{-0.25}^{+0.24}$& $>$--0.10\\[5pt]   
J163950+434003  &  16:39:50.52  & +43:40:03.7   &  3.990 &  3.668  & 3 &  600, 450                   &17.80--19.10& --2.52$_{-0.15}^{+0.18}$& --1.99$\pm0.10$& $>$0.55\\
\hline
\label{table:ESI}
\end{tabular}
\begin{tablenotes}
\item[a] Denotes a partial LLS.
\item[b] Observed during $18^{\circ}$ twilight.
\end{tablenotes}
\end{threeparttable}
\end{table*}

\section{Observations and Data Reduction}\label{sec.obs}

Our sample of 33 LLSs is a subset of the 194 LLSs with $\zlls\geq3.3$ and $\nhi\geq17.5\cm{-2}$ found in SDSS DR7 by \citet{Prochaska}, the same parent sample as \citetalias{Cooper}. However, unlike the ``metal-poor'' sample from \citetalias{Cooper}, they were not further screened for (lack of) metal absorption. Instead, they were grouped into three ``tiers'' based on the prominence of their metal absorption lines upon visual inspection of the SDSS spectra. The tiers are classified as no metals (Tier 1; 27\% of the 194), possible metals (2; 15\%), and obvious metals (3; 58\%). 

The \citeauthor{Prochaska} catalog includes DLAs, which we are not interested in here because LLS and DLA populations have different metallicity distributions (e.g., \citetalias{Cooper}). DLAs can be metal poor \citep{2015ApJ...800...12C}, but still result in metal-line absorption due to the sheer amount of gas; hence, most DLA candidates would be in our Tier 3. We visually inspected the Tier 3 SDSS spectra that are not in the Keck Observatory Archive (KOA) for damping wings. Then, we excluded follow-up on 34 of the 87 non-KOA Tier 3 LLSs likely to be DLAs. Thus, unlike \citetalias{Cooper}, which only selects from Tier 1, we selected from all tiers, ultimately: 10 Tier 1, 10 Tier 2, and 13 Tier 3. To preserve uniformity in the the spectra included in our sample, we opted not to use LLSs already present in the KOA in this paper, as they have an assortment of spectral resolutions and data qualities.  Thus, we had a stratified sample from which to gauge the full extent of LLS metallicities.\footnote{In KOA, there is a mix of tiers (8 Tier 1, 4 Tier 2, and 26 Tier 3) and spectrographs (ESI, HIRES). We ultimately decided to proceed with our homogeneously constructed sample, which is also why we do not incorporate the 17 LLSs from \citetalias{Cooper}.} Although the stratified sample may exhibit some biases and hence not be completely representative, it ensures that we include a range of LLSs that can still be used to make general statements about the metal distribution function. 

To determine whether the three tiers are consistent with being drawn from the same parent metallicity distribution, Kolmogorov-Smirnov (KS) tests were performed on cumulative distribution functions (CDFs, see Section \ref{sec:pdf}) constructed for each tier independently. In all three comparisons, we found no statistical difference to indicate distinct parent populations for the tiers; note this is not evidence they are from the same parent population.

We observed the quasars toward which these 33 LLSs were identified, using the Keck Echellette Spectrograph and Imager \citep[ESI,][]{ESI} on UT 17--18 January and UT 19 April 2015 using 0.75\arcsec\ slits. ESI covers the optical spectrum from 0.39--1.1\,microns and, with 0.75\arcsec\ slits, has a resolution of (full-width at half-maximum) ${\rm FWHM} \approx 50\kms$. (SDSS spectra have  ${\rm FWHM}\approx150\kms$.) Our mean redshift $\overline\zlls=3.7$ corresponds to an observed wavelength of 4270\,\AA\ for the Lyman break. Observational details are listed in Table \ref{table:ESI}.  We processed the raw frames into 1D, flux-calibrated spectra using the {\tt XIDL}\footnote{See \url{http://www.ucolick.org/~xavier/IDL/}.} software package.

We confirmed the redshift of each LLS (found in the SDSS survey by matching the observed break) with Lyman series lines visible in our higher-resolution spectra and also with metal lines where available (typically either \ion{Si}{4} $\lambda\lambda1393,1402$ and\slash or \ion{C}{4} $\lambda\lambda1548,1550$). For all sightlines, we selected the highest redshift LLS. Typically, the redshifts matched at the $\vert \Delta z \vert \approx 0.001$ level. Two systems had atypical discrepancies, both with $\vert \Delta z \vert \approx 0.15$: J144144+472003 and J122027+261903. For both these sightlines, our quoted \zlls\ was higher than that measured from the Lyman break in SDSS spectra. J144144+472003 contains several partial LLSs (pLLSs). (We label as pLLSs those systems that had only a partial Lyman break in their spectra, allowing more precise measurements for $\log\nhi\lesssim 17.6$.) The Lyman break located at $z=3.443$ in the SDSS appears to be due to several pLLSs at different redshifts; we include the highest-redshift pLLS in our sample. J122027+261903 has a higher-redshift LLS that is not apparent in SDSS. Since the lower-redshift system has an unreliable \nhi\ measurement due to absorption by the higher-redshift LLS, we only analyze the higher-redshift LLS.  The data quality is fairly uniform; most spectra have a signal-to-noise ratio of 20--30 at the wavelength of \ion{C}{4} at the LLS redshift.

\begin{figure*}[tbh]
\centering
\plotone{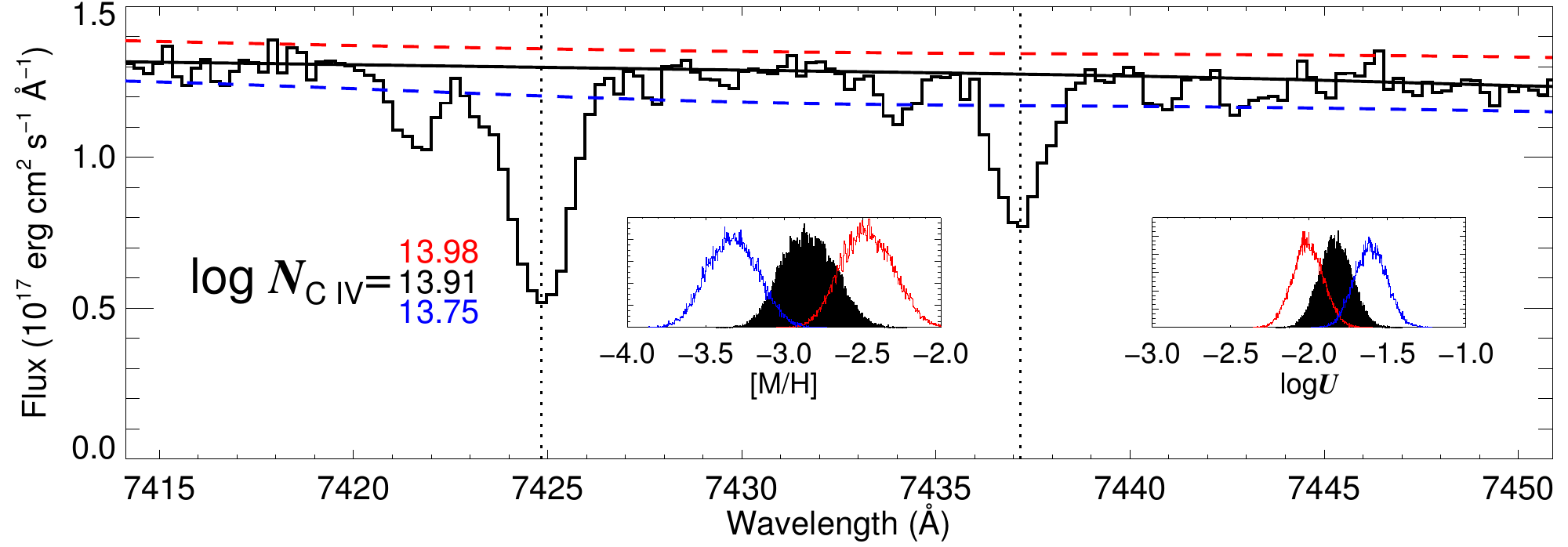}
       \caption[highlowfit]{Continuum fit (black curve) to the spectrum of J140248+014634 (black histogram) around the $\zlls = 3.796$ \ion{C}{4} doublet. Also shown are unrealistically high (red) and low (blue) continuum fits used to gauge how uncertainty in continuum placement influences ionization modeling posteriors.  Insets show the resulting posterior metallicity \meh\ and ionization parameter \logu\ distributions with nominal, high, and low fits to the continuum around all measured ions. Although the \ion{C}{4} column density only changes by about 0.1 dex, the distribution shifts by about 0.5 dex with the offset continuum placements because ionic column densities based on weaker lines (e.g., \ion{Si}{2} $\lambda1526$) are more sensitive to continuum placement. Varying the continuum fit within more reasonable bounds only changes the resulting metallicities by about 0.1 dex.}
     \label{fig:highlowfit}
\end{figure*}

We estimated and normalized the continuum level of each quasar spectrum interactively, using a cubic-spline interpolation fit. To determine the effects of continuum placement on our derived metallicities, we ran tests using extreme values of the continuum fit. The J140248+014634 spectrum is shown in Figure \ref{fig:highlowfit} with an extremely high and low continuum placement around the \ion{C}{4} doublet. The posterior metallicity distribution corresponding to each continuum placement was found using the methods discussed below. For this system, the extreme fits led to a $\sim\!0.5\,$dex change in the posterior metallicity and a $\sim\!0.2$--0.3\,dex change in \logu. More realistic continuum placement leads to uncertainties that are lower than the statistical uncertainties from our modeling. Moreover, some systems are more robust to continuum placement and even the extreme continuum fits are within the modeling uncertainty.  Thus, we do not quantify continuum placement uncertainties since they do not significantly contribute to our overall error budget.

\section{Analysis}\label{sec.analysis} 

The analysis methods used in this work are very similar to those presented in \citetalias{Cooper}, and we refer the reader to that work for more details.  Here we summarize the main steps and highlight minor changes incorporated since publication of \citetalias{Cooper}.

\subsection{\ion{H}{1} and Metal Ion Column Density Measurements\label{subsec.nhi}}

We manually determined a range of possible \ion{H}{1} column densities for each LLS, using an adapted version of the {\tt x\_fitdla} routine in {\tt XIDL}.  This GUI interactively overplots Voigt profiles of tunable redshift, Doppler parameter $b$, and \nhi\ on the spectral data for the user to estimate upper and lower bounds on \nhi.

\begin{deluxetable}{lllllll}
\tablecaption{Metal Column Densities \label{table:metals}}
\tablehead{\colhead{Ion}&\colhead{$\lambda_{\textrm{rest}}$ (\AA)}&\colhead{$\log \N{AODM}$}&\colhead{$\log\N{adpt}$\tablenotemark{a}}&\colhead{$\log \N{pred}$\tablenotemark{b}}}
\startdata

\cutinhead{J011351--093551\hspace{.5 cm}$\zlls=3.617\tablenotemark{c}$\hspace{.5 cm}$\log\nhi=17.80$--19.10}
\ion{Al}{2}&1670&   $12.56\pm0.02$    &   $12.56\pm0.02$ &   $12.87 \pm 0.09$\\
\ion{Al}{3}&1854&   \nodata         &   \nodata &   $12.57 \pm 0.07$\\
\ion{Al}{3}&1862&   $12.72\pm0.08$    &   $12.72\pm0.08$ &\nodata    \\
\ion{C}{2}&1334&    $14.15\pm0.01$    &   $14.15\pm0.01$ &   $14.10 \pm 0.08$\\
\ion{C}{3}&\nodata&\nodata&\nodata&$15.42 \pm 0.06$\\
\ion{C}{4}&1548&    $14.32\pm0.01$    &   $14.33\pm0.004$    &   $14.27 \pm 0.08$\\
\ion{C}{4}&1550&    $14.36\pm0.01$    &   \nodata &\nodata    \\
\ion{Si}{2}&1304&   $13.49\pm0.07$    &   $13.49\pm0.07$ &   $13.42 \pm 0.07$\\
\ion{Si}{3}&\nodata&\nodata&\nodata&$14.42 \pm 0.06$\\
\ion{Si}{4}&1393&   $13.76\pm0.01$    &   $13.76\pm0.01$ &   $ 13.95 \pm 0.06$\\
\cutinhead{J034402--065300\hspace{.5 cm}$\zlls=3.843$\hspace{.5 cm}$\log\nhi=17.80$--19.40}
\ion{Al}{2}&1670&   $<11.80$  &       $<11.80$  &   $11.93 \pm 0.12$\\
\ion{Al}{3}&1854&   $<12.25$  &       $<12.25$  &   $11.65 \pm 0.10$\\
\ion{Al}{3}&1862&   $<12.54$  &       \nodata &   \nodata\\
\ion{C}{2}&1334&    $13.33\pm0.07$    &   $13.33\pm0.07$ &   $13.14 \pm 0.08$\\
\ion{C}{3}&\nodata&\nodata&\nodata&$14.49 \pm 0.07$\\
\ion{C}{4}&1548&    $13.53\pm0.02$    &   $13.54\pm0.02$    &   $13.31 \pm 0.08$\\
\ion{C}{4}&1550&    $13.57\pm0.04$    &   \nodata &   \nodata\\
\ion{Si}{2}&1260&   $12.28\pm0.07$    &   $12.28\pm0.07$ &   $12.41 \pm 0.08$\\
\ion{Si}{3}&\nodata&\nodata&\nodata&$13.50 \pm 0.09$\\
\ion{Si}{4}&1393&   $12.69\pm0.06$    &   $12.73\pm0.05$ &   $13.02 \pm 0.06$\\
\ion{Si}{4}&1402&   $12.83\pm0.09$    &   \nodata &   \nodata\\

\enddata
\tablenotetext{a}{Adopted column densities. For saturated lines, we use lower limits. For non-detections, we use 3-$\sigma$ upper limits.}
\tablenotetext{b}{Column density as predicted by the {\tt Cloudy} model using the ionization and metallicity parameters obtained via MCMC modeling.} 
\tablenotetext{c}{Errors to the redshift were generally on the order of 10$^{-3}$.}
\tablecomments{This table is published in its entirety in the electronic edition; a portion is shown here as an example.}
\end{deluxetable}
 \

As the \ion{H}{1} absorption lines were all saturated, their measured column densities were highly uncertain. In most cases, the lower limit on \nhi~ was established by the existence of the full Lyman break, which only occurs when $\log\nhi>17.8$.  Two systems (J131453+080456 and J144335+334859) showed weak Ly$\alpha$ damping wings and hence had reliably larger column densities. Two others (J144144+472003 and J121058+182119), classified in SDSS spectra as LLSs \citep{Prochaska}, were revealed as pLLSs with ESI, permitting a low but highly accurate determination of \nhi.  For the remaining systems, upper bounds on \nhi\ were determined by increasing the column density until absorption at one or more transitions fell below the data for a reasonable Doppler parameter (typically 20\kms). Examples and further details are given in \citetalias{Cooper} (Section 3.1).

Within the range of upper and lower bounds, we treat all values of the \ion{H}{1} column density as equally likely (i.e., a flat prior). In \citetalias{Cooper} we showed that the uncertainty introduced to metallicity measurements by this assumption is comparable to uncertainties from ionization modeling at a single \nhi, primarily because the total hydrogen column density \nh\ changes by $\sim$0.3\,dex for a change of 2\,dex in neutral \nhi. The {\tt Cloudy} runs used in \citetalias{Cooper} (Figure 7) show that the \ion{C}{4} and \ion{Si}{4} column densities are robust, changing by $\sim$0.2\,dex over a 2\,dex change in \nhi~(holding metallicity and density fixed), while \ion{C}{2} and \ion{Si}{2} vary by $\sim$0.8\,dex over the same neutral hydrogen interval.

For each LLS, we examined the absorption lines from the ionic species: \ion{Si}{2} $\lambda1260$, $\lambda1304$, and $\lambda1526$; \ion{C}{2} $\lambda1334$; \ion{Si}{4}; \ion{C}{4};
\ion{Al}{2} $\lambda1670$; and \ion{Al}{3} $\lambda\lambda1854,1862$. Several other commonly studied transitions (e.g., \ion{O}{1} $\lambda1302$ and \ion{Fe}{2} $\lambda1608$) fall within the wavelength range of our spectra, but due to a combination of weak oscillator strengths and small column densities, they only yield very large column density upper limits for these absorbers (i.e., the data do not constrain their column densities to within relevant values) and are not included in our analysis. For each line that was not obscured by noise, interloping absorption, or the Ly$\alpha$ forest, the column density (or a limit) was measured using the apparent optical depth method \citep[AODM;][]{Savage}. For each absorber, we manually assign a velocity width for the AOD measurement, based on the absorption features. We found the 3-$\sigma$ upper bounds for lines without any absorption detected using a Monte Carlo technique. We added Gaussian noise to each pixel according to the error spectrum to determine how large the column densities could be while still showing no observable absorption. Through many realizations of this process, we constructed a column density distribution. The 3-$\sigma$ upper limit column density was then chosen as the column density that was larger than 99.7\% of the distribution. These limits are close to those found with simpler AOD procedures and were initially used to gauge whether mismatching Al II limits (see below) could be due to inaccurate upper-limits.

For species with multiple absorption lines, we compare AOD profiles to test for saturation and assign lower limits to saturated lines. We perform a $\sigma$-weighted average of measured column densities of unsaturated lines. We also employed a saturation test for species with only one line available. In these cases, we created many multi-component absorption models of various velocity structures and found best-fit column densities for each model. If more than 5\% of the column densities were larger than initially measured, it would show that the absorption line may be saturated. Using this process, we found no convincing cases for this type of single line absorption in our ESI spectra. In Table \ref{table:metals}, the measured metal column densities for each system are listed.

While the lack of saturation may come as a surprise, there are numerous examples in the literature of high-redshift LLSs (and even some super-LLSs/sub-DLAs) observed at higher spectral resolution in which the lines we use are unsaturated \citep{ 2005A&A...440..819R,2010MNRAS.409..269S,2010ApJ...708.1221P,2014A&A...572A.102F,2015ApJS..221....2P}. At these redshifts, optically thick systems showing saturation in the lines we use are typically at higher \nhi\ \citep[e.g.,][]{2006ApJ...648L..97P} and/or have unusually high metallicities \citep{Crighton}. We also note that we do not consider many strong lines that may be saturated but fall within the Lyman-$\alpha$ forest (e.g., \ion{Si}{3} $\lambda1206$) or are redshifted beyond our spectral range (e.g., \ion{Mg}{2} $\lambda\lambda 2796,2803$).

As an appendix (Figure \ref{fig:stack}), we include portions of each of the LLS spectra in our sample, normalized and extracted around several Lyman series transitions and the metal lines used. 

\subsection{Ionization and Metallicity Modeling}\label{subsec.cloudy}

Detailed ionization modeling is required to extract estimates of heavy-element abundances from the column density measurements described.

First, we construct a grid of ionization models using {\tt Cloudy} \citep[version c13.02, last described by][]{Ferland}. {\tt Cloudy} calculates the temperature and ionization of diffuse interstellar/intergalactic gas for the inputs of \ion{H}{1} column density (\nhi), metallicity (\meh), and ionization parameter (\logu). For any combination of these input parameters, {\tt Cloudy}  outputs ionization fractions for the specified elements and associated metal-line column densities.

We assume a geometry of a large, uniform gas slab with solar relative abundances \citep{Asplund}. The slab thickness is determined dynamically by {\tt Cloudy}  to match the input \ion{H}{1} column density. The ionization parameter is the ratio of the volume density of photons able to ionize neutral hydrogen to the volume density of hydrogen $n_{\rm H}$, defined as: 
 
\begin{equation} 
U=\frac{\displaystyle 4\pi\int_{\nu_{\text{LL}}}^{\infty}
  \frac{J_{\nu}}{h\nu} d\nu }{n_{\rm H}c},  \label{eqn.U}
\end{equation} 

where $J_\nu$ is the spectral flux of the ionizing background; $h\nu$ is the energy of a photon with frequency $\nu$;  $c$ is the speed of light; and $\nu_{\rm LL}$ is the Lyman limit (1\,Ryd). The ionization parameter implicitly accounts  for gas density. Gas in the LLS is ionized by the ambient background radiation field, which contains integrated contributions from galaxies and QSOs, as well as an account of \ion{He}{2}\ Ly$\alpha$ absorption from the IGM\citep{haardtmadau}.

\begin{figure}
  \begin{center}
\includegraphics[width=8.9cm]{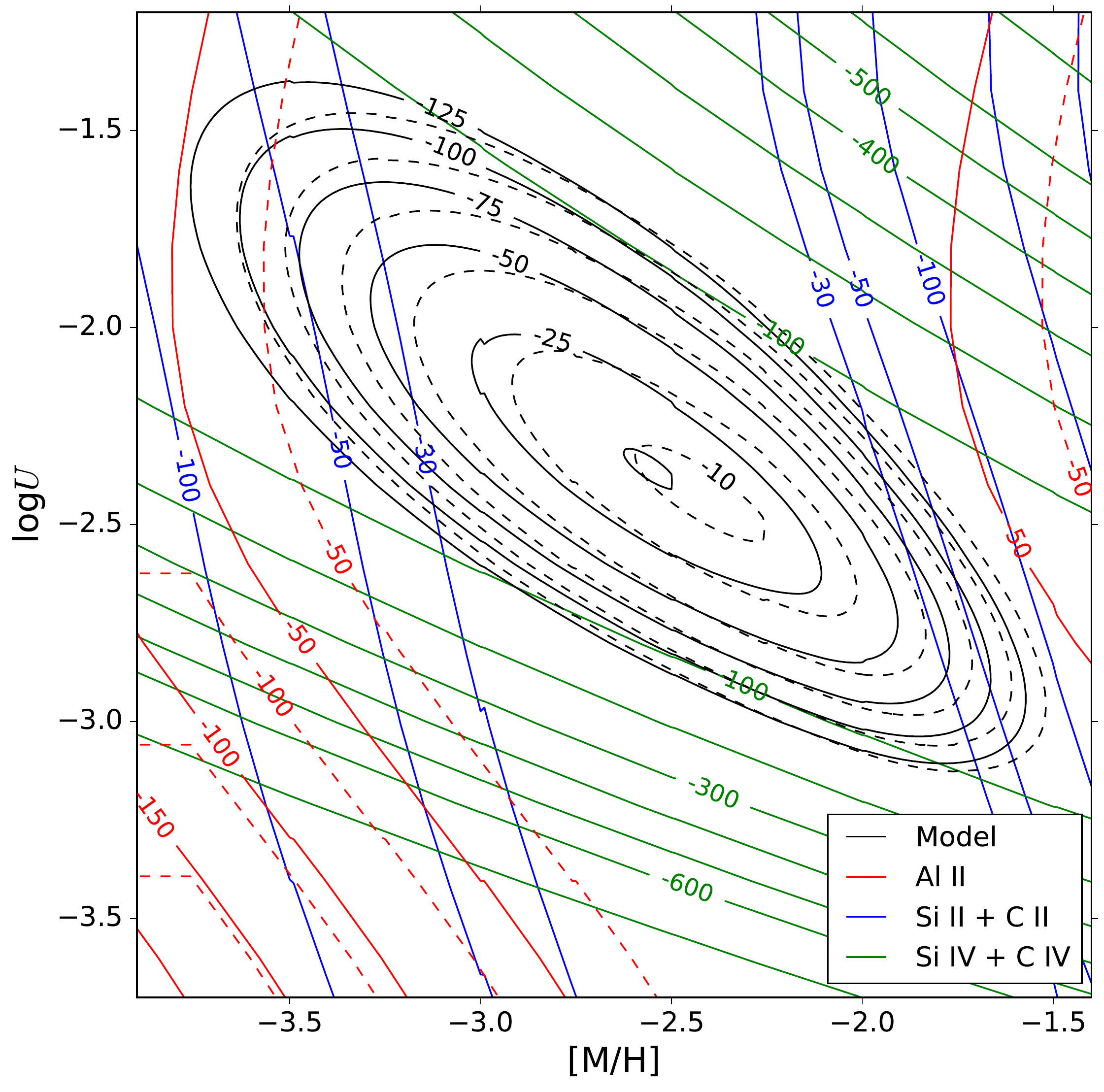}
       \caption[lnlike]
       	{Contours showing the contributions to the log-likelihood function from different ionic species (for a fixed \nhi), for the LLS along the sightline to J141831+444937. Increasingly negative values correspond to a less likely region of parameter space. The black contours correspond to the full model using all ions, and the colored contours only use the ions given in the legend. Dashed curves assume that aluminum is at the same solar-relative abundance as the other metals, whereas the solid curves take the central value of \aratio\ for this LLS ($\aratio=0.24$). Singly ionized silicon and carbon (blue) constrain the metallicity, while the triply ionized species constrain the ionization parameter/density. Requiring a solar-relative aluminum abundance forces the models to higher metallicity and density. (The discontinuity in the red dashed \ion{Al}{2} contours is simply where $\meh-\aratio<-4$, the edge of our grid.)}
     \label{fig:lnlike}
  \end{center}
\end{figure}

Our {\tt Cloudy} grid spans \meh~and \logu~from $[-4,-1]$ and $[-3.8,-1]$, respectively, in steps of 0.1\,dex. We found that this choice covered the range of parameters measured in both subsamples in \citetalias{Cooper}, and all of the LLSs analyzed in this paper appear to fit well inside this range. The grid also has a $\log\nhi$ step size of 0.1\,dex and a redshift step size of 0.1. We use this grid to define an interpolating function, \N{model}, for each ionic column density in (\meh, \logu, \nhi, $z$) space.

Given the observed column densities (\N{adpt}) and a {\tt Cloudy}-based forward model of column densities spanning our physical parameter domain, we create the likelihood function:
\begin{equation}
\mathcal{L}=\prod 
\exp\bigg[{-\frac{(\N{adpt}-\N{model})^{2}}{2\sigma_{\N{adpt}}^{2}}}\bigg], \label{eqn.lnL}  
\end{equation} 
which assumes Gaussian statistics. This product is taken over all ions constraining the absorber in question, and $\sigma_{\N{adpt}}$ represents the error in the measured column density for that ion. For detections, a Gaussian is used to describe the likelihood function for each ion. For upper and lower limits, a one-sided Gaussian is applied. In practice, this is implemented by setting $\mathcal{L}=1$ if the model column density falls below a measured upper limit (or above a lower limit), while letting $\mathcal{L}$ drop off along a Gaussian probability density function (PDF) if it violates the measured limit. The natural log of this likelihood function is used to avoid computation instabilities.

To see the constraints imposed on the likelihood function from individually measured ions, it is instructive to examine raw likelihood contours. Figure \ref{fig:lnlike} shows one such example of J141831+444937 in which we isolated the contribution of singly ionized species (\ion{Al}{2} and the combined contours of \ion{Si}{2} and \ion{C}{2}, which are very similar and have been combined for clarity) to $\ln\mathcal{L}$ from that of the triply ionized species (\ion{Si}{4} and \ion{C}{4}, combined).  It is clear from the figure that \meh\ is primarily constrained by the singly ionized species, with the triply ionized species primarily discriminating \logu. The black ``bull's-eye'' shows joint likelihood contours around the solution of the model that uses \ion{Al}{2}, \ion{Si}{2}, \ion{C}{2}, \ion{Si}{4}, and \ion{C}{4}. The likelihood contours fit well in the parameter space of our model, evidencing that the ranges selected for our parameters are large enough.

With a likelihood function in hand to measure the model's goodness-of-fit for each point in (\meh, \logu, \nhi) space (at the redshift of each LLS), we explored this space using an MCMC simulation, implemented with the open-source \texttt{Python} package {\tt emcee} \citep{Foreman}.  We assumed flat priors for \nhi, with a range for each LLS assigned using the manual Voigt profile plausibility fitting described in Section \ref{subsec.nhi}. We exclude the first 100 steps taken by the MCMC ``walkers'' as a ``burn-in'' phase so that their starting locations do not bias the result. We took our metallicities and \logu\ values to be the median of the walker results, with 1-$\sigma$ errors given by the 68.3\% confidence interval.

Following the procedure of \cite{Crighton}, we also forced a minimum uncertainty of 0.1\,dex on the adopted column densities because assumptions of uniformity and equilibrium in our modeling likely do not capture the full description of the gas. In agreement with \citeauthor{Crighton}, we also found that setting the minimum uncertainty to 0.15 or 0.05\,dex hardly changed our results, supporting the choice of 0.1\,dex as reasonable. Most importantly, as described by \citeauthor{Crighton}, this allows a larger range of solution space to be explored by the MCMC walkers as an unreasonably small error in one ionic transition (or overconfidence in the fidelity of the {\tt Cloudy} model or relative abundance for this transition) might otherwise over-constrain the global solution. 

Typically, the MCMC walkers converged on a location in \meh-\logu\ space, similar to that outlined by the black contours in Figure \ref{fig:lnlike} (see also Figure \ref{fig:mesh}). It is perhaps surprising that in several cases our posteriors did not converge, but rather bifurcated into two possible parameter regions, having similar (yet poor) qualities of fit.  Deeper examination revealed that this situation was particular to systems where \ion{Al}{2} and/or \ion{Al}{3} were included as an empirical  constraint in the MCMC simulation.  In all such cases, the observed aluminum column densities (or upper limits) were lower than the relevant model expectations based on solar relative abundances.

\begin{figure*}
  \begin{center}
    \leavevmode
\includegraphics[width=0.49\textwidth]{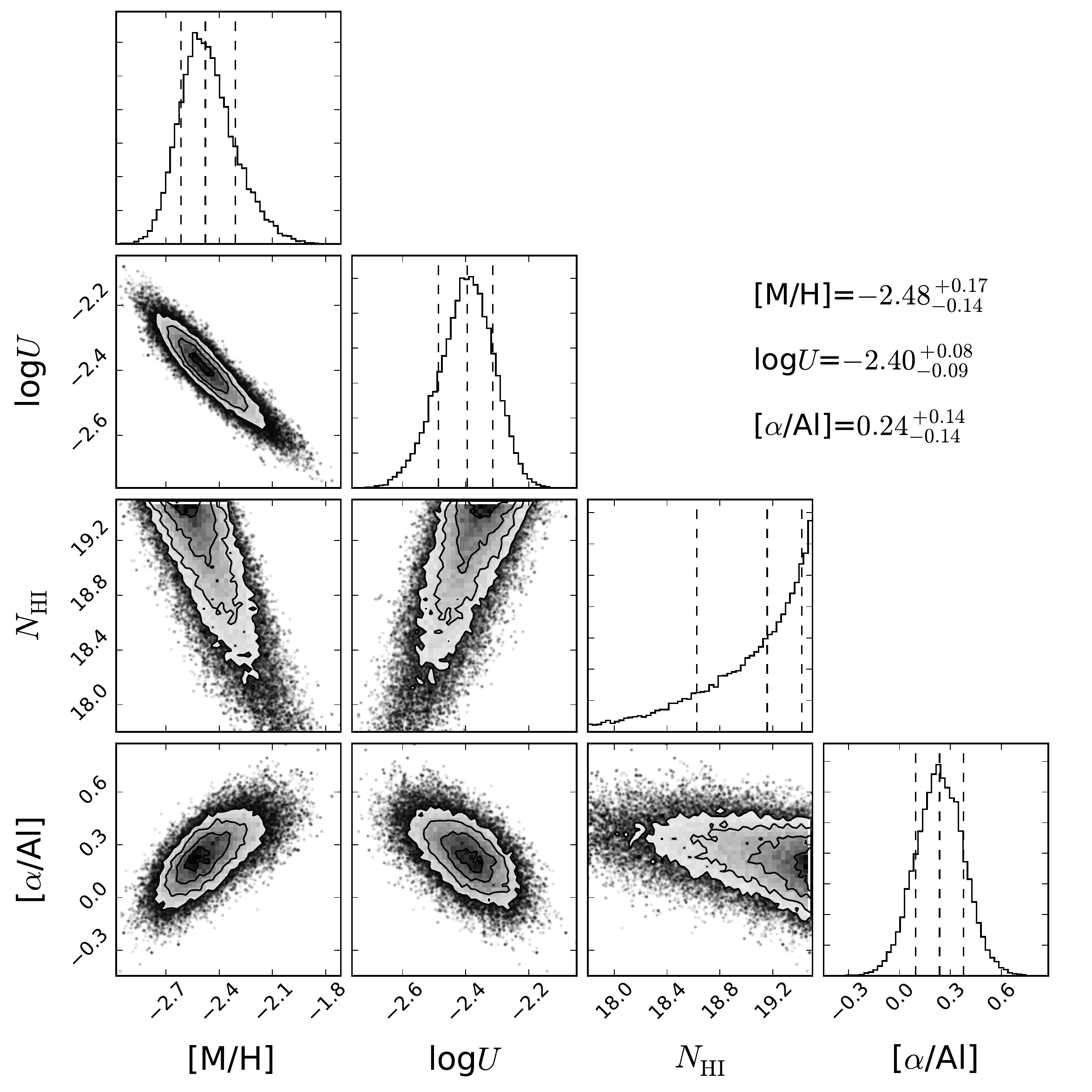}
 \includegraphics[width=0.49\textwidth]{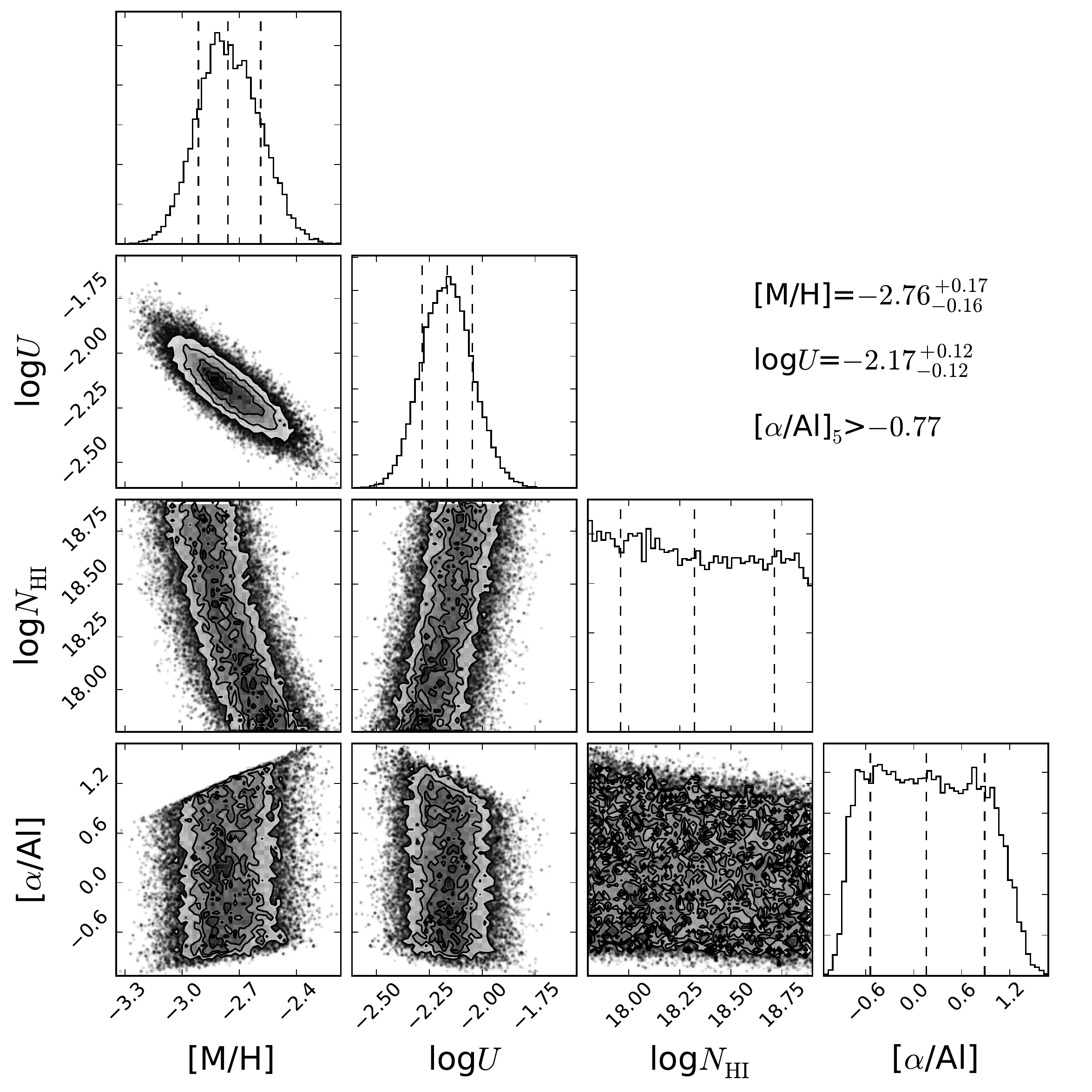}\\
\includegraphics[width=0.49\textwidth]{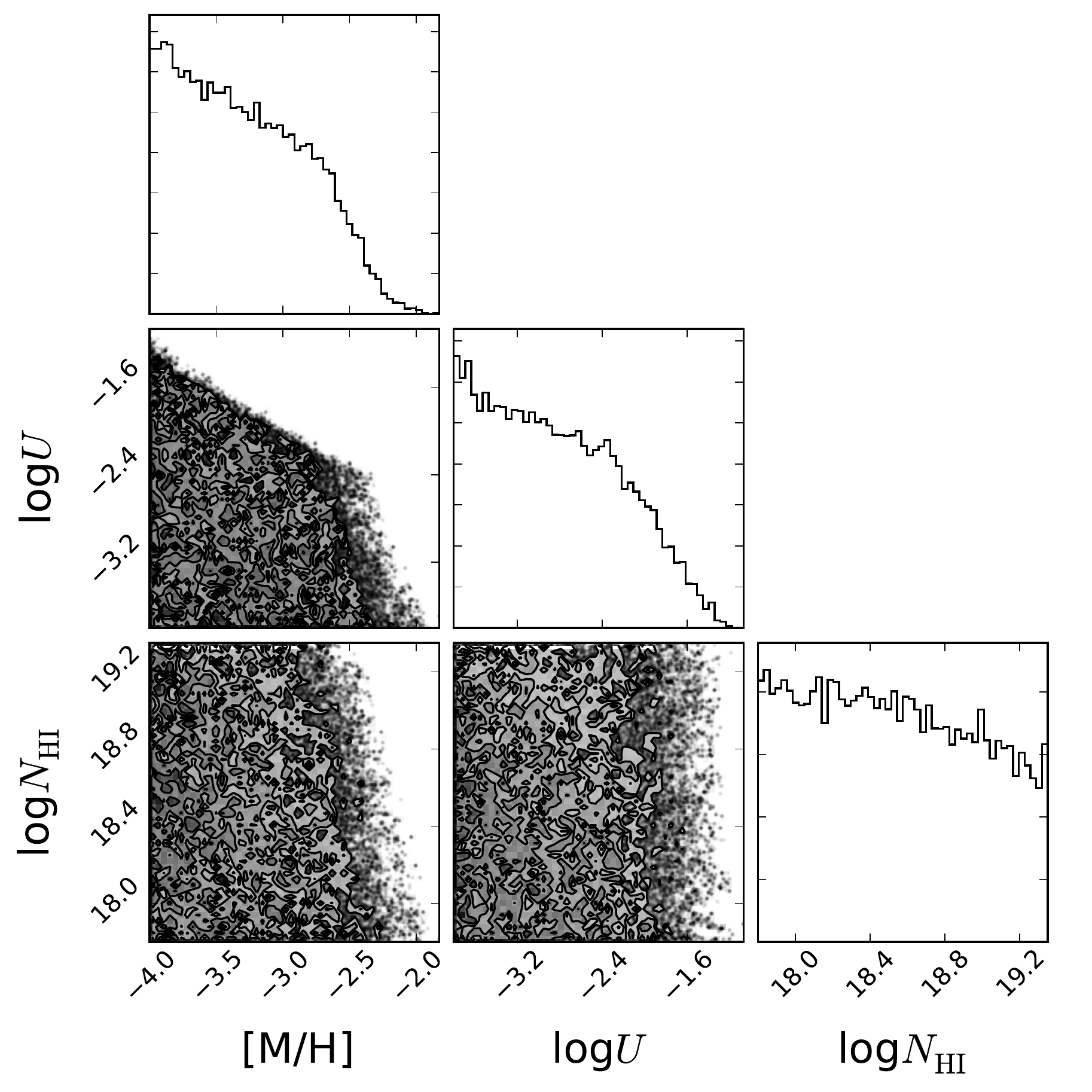}
 \includegraphics[width=0.49\textwidth]{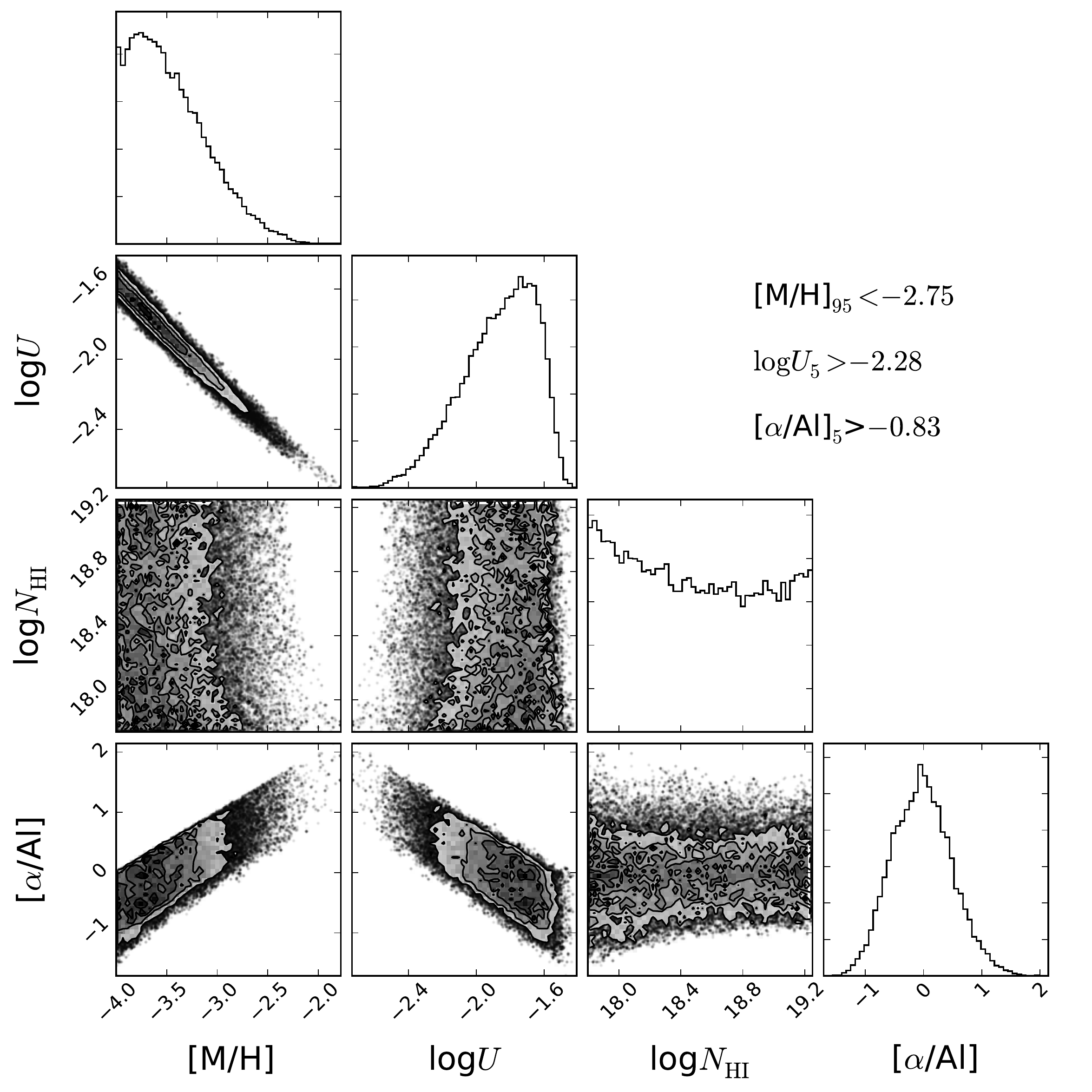}
       \caption[Triangle]{Four examples of  MCMC modeling posterior distributions. 
         The contours of the walkers' likelihood space show the
         pairwise relations between \meh, \logu, $\log\nhi$, and
         \aratio~(when applicable). Histograms for each
         of these variables are also given. \textit{Top Left:} 
         J141831+444937: example of a detection, where values for
         \meh, \logu, and \aratio~can be determined from the metal
         absorption lines.   \textit{Top Right:}
         J122000+254230: example of a detection for \meh~and \logu,
         while \aratio~remains a lower limit. \textit{Bottom Left:} J103048+391234:
         example of a ``Type 1'' upper limit for \meh~as 
         (see Sections \ref{subsec.cloudy} and \ref{sec.results}). In these cases, an MCMC simulation is not used to
         determine the upper limit of \meh, and the values for \meh~and
         \logu~shown in the figure are not used. Instead, they are
         treated as discussed in  Section \ref{subsec.cloudy}. \textit{Bottom Right:}
         J082340+342753: example of a ``Type 2'' upper limit for \meh~and
         lower limits for \logu~and \aratio. 
         The number (fraction) of the LLSs that each panel applies to is as follows (left-to-right, top-to-bottom): 9 (27\%), 8 (24\%), 14 (42\%), and 2 (6\%). 
    The {\tt Python} module
         \texttt{Corner} \citep{Foreman} was used to format the  MCMC figures. 
       }
     \label{fig:triangle}
  \end{center}
\end{figure*}

In systems of higher \nhi, a paucity of a refractory element such as aluminum might be taken as evidence of dust depletion \citep[e.g.,][]{Prochaska2002}.  We do not see such a pattern in carbon and silicon, so it is possible that the discrepancy could have a nucleosynthetic origin (aluminum is not an $\alpha$-element). An interesting future test would be to observe the aluminum-depleted systems in the infrared to include \ion{Fe}{2} and \ion{Mg}{2} as additional model constraints.  For the present paper, we simply modify our MCMC procedure, introducing an \aratio~variable.  Aluminum abundances are then drawn from a model with $\xh{Al}=\meh-\aratio$ with a flat prior on \xh{Al} between $-4$ and $-1$. This produced much better agreement with the data, at the obvious cost of introducing another free model parameter. Further discussion of aluminum abundances and possible interpretations is deferred to Section \ref{sec:arat}. A similar discrepancy, in which aluminum was depleted by 0.3 dex relative to $\alpha$-elements, was noted by \citet{Crighton2013} in a single component of an absorber.

Figure \ref{fig:lnlike} demonstrates the effect of this modification. The solid contours show the likelihood function for a particular system taken at the best-fit value of $\aratio=0.24$ found by the 4D MCMC, whereas the dashed contours have $\aratio=0$ (i.e., solar relative abundance). The MCMC walkers are mostly within the $\mathcal{L}=-25$ (black) contour. In the model with $\aratio=0$, aluminum column densities drive the solution to higher metallicity and lower ionization parameter, forcing the model into tension with the other measured ions. This effect is more pronounced for systems with an \ion{Al}{2} non-detection, as the likelihood (with \aratio\ fixed at zero) quickly falls off for models predicting \ion{Al}{2} column densities inconsistent with the upper-limit derived from the spectrum.

As useful as the likelihood contours are to visualize how each metal line contributes to the model's solution, they are still limited as they only show a cross section of the solution space in the 2D plane of \logu~and \meh. The third dimension of \nhi~cannot be shown in this way and is in fact fixed to an intermediate value in order to produce these figures. However, the entire 4D parameter space is probed by the MCMC simulation, and our posterior distributions reflect this full projection in Figure \ref{fig:triangle}.

Four examples of the MCMC-determined walker space in 4D (or 3D where applicable) are shown in Figure \ref{fig:triangle}. Unlike in Figure \ref{fig:lnlike}, Figure \ref{fig:triangle} is able to show the best-fit walker solution space for each of the pairs between \meh, \logu, \nhi, and \aratio. 

Figure \ref{fig:triangle} features an example of each type of measured value or limit possible for \meh, \logu, \nhi, and \aratio. The top left panel shows the simplest case, where a definite value can be determined for \meh, \logu, and \aratio. The top right panel shows an example of a best-fit value for \meh~and \logu~when only a lower limit for \aratio~can be evaluated.  Such systems typically arise in cases where \ion{Al}{2} is a non-detection.

The bottom left panel displays results for the situation where no metal lines are detected, the so-called ``Type 1'' upper limit of \citetalias{Cooper}. In these cases, the MCMC walkers cannot converge on a solution. The data cannot constrain \logu, and at each value of \logu\ we obtain an upper limit to the metallicity, resulting in an allowed region in \meh-\logu\ space. The metallicity is generally constrained by the \ion{C}{2} and \ion{Si}{2} column density upper limits at low \logu\ and by \ion{C}{4} and \ion{Si}{4} upper limits at high \logu. Larger values of \logu\ correspond to {\em lower} metallicity upper limits (see Figure~\ref{fig:mesh}). When a single value for \meh\ is needed for these LLSs, the most conservative (i.e., largest) upper bound to \meh\ is found by (i) setting \logu\ to a conservative value of $\logulim=-3$, (ii) setting \nhi\ equal to the minimum \nhi\ possible for the system (typically $10^{17.8}\cm{-2}$, a conservative assumption), and (iii) determining the uppermost value of \meh\ that can be used without violating the observed 3-$\sigma$ column density limits. We exclude aluminum because its depletion relative to the $\alpha$--elements would (falsely) lead to more aggressive (i.e., smaller) upper limits. The value of \logulim\ is justified in \cite{Fumagalli} based on the relatively small number (until recently) of LLS ionization measurements at comparable redshift and is extremely conservative as it is far below all of our measured values.\footnote{Our lowest measured ionization parameter is \logu=$-2.46$, along the sightline to J083941+031817.} Quantifying how conservative these limits are in terms of uncertainties is not straightforward, but we note that they are likely more than 3-$\sigma$ because they are derived from 3-$\sigma$ column density upper limits at extremal values of \logu\ and \nhi. 

The bottom right panel exemplifies a configuration where the data provide (i) an upper limit on \meh\ based on non-detections of \ion{C}{2} and \ion{Si}{2}  but (ii) a lower limit on \logu~based on the detections of \ion{Si}{4} and/or \ion{C}{4}.  These systems, named ``Type 2'' limits in \citetalias{Cooper}, display clear degeneracy between \meh ~and \logu.  We assign upper bounds to \meh\ for these absorbers according to the metallicity that 95\% of the walker steps are below. This type of limit allows us to better constrain both the metallicity and ionization parameter than ``Type 1'' limits. These systems generally result in negative lower limits to \aratio\ (i.e., they are unconstraining).

\section{Results}\label{sec.results}

Before describing our sample's \meh, \logu, and \aratio\ and discussing trends, we succinctly summarize our terminology.

As described in Section \ref{sec.obs}, we categorized the \citet{Prochaska} LLSs into three ``tiers'' prior to ESI follow-up, based on the SDSS spectra: 1---no metal lines; 2---possible metal lines; and 3---likely metal lines. 

Following \citetalias{Cooper}, we have two ``types'' of \meh\ upper limits: 1---no metal lines in ESI and, as needed, adopted \meh\ upper limits conservatively taken at $\logulim=-3$; and 2---no low ion lines so \logu\ is not constrained and \meh\ upper limit is defined at a value above 95\% of MCMC steps.  

For the Tier 1 absorbers, we had six heavy-element detections at ESI resolution (${\rm FWHM} \approx 50\kms$), including one pLLS, and four Type 1 upper limits. The Tier 2  sample had seven detections, including one pLLS, two Type 1 upper limits, and one Type 2 upper limit. The Tier 3 sample had 11 detections, one Type 1 upper limit,\footnote{This sightline was classified as Tier 3 due to an interloping absorption line that we misidentified as \ion{C}{2} in the SDSS spectrum.} and one Type 2.

\subsection{Metallicity and Ionization of LLSs}

\begin{figure*}
  \begin{center}
    \leavevmode
\includegraphics[width=14cm]{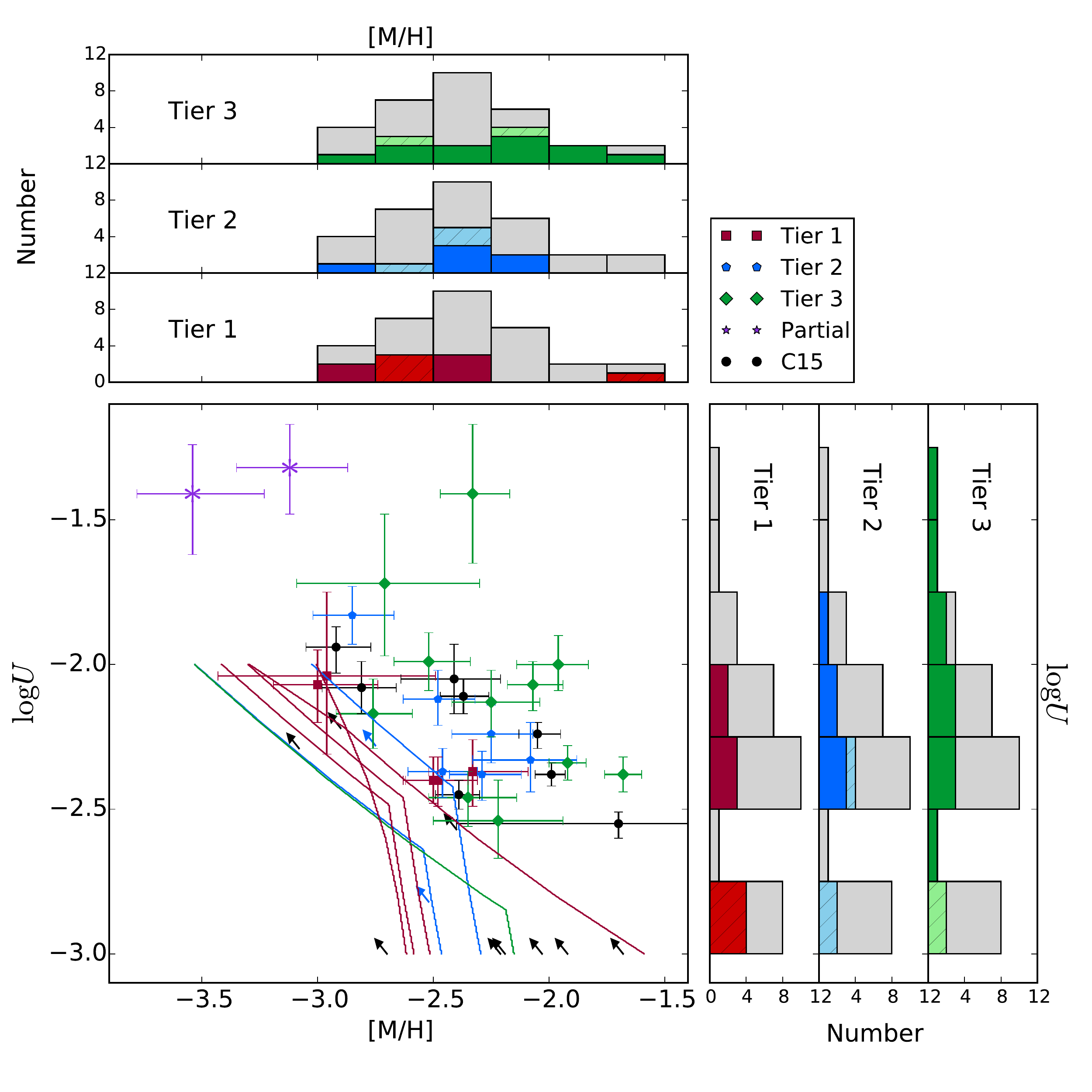}
\caption[mesh]{Scatter plot showing \logu\ vs. \meh\ for the LLSs with error bars from the MCMC posterior distributions. Markers with error bars show systems with definite best-values for \logu\ and \meh, while the lines trace the upper bound of the Type 1 limit cases (no metal lines) and the arrows show Type 2 limits (no low-ionization lines). The tiers are indicated by different colors: Tier 1 (defined as no metals in SDSS) is red; Tier 2 (possible metals of low SNR) is blue; and Tier 3 (likely SDSS metals) is green. In black are the detections and limits from the metal-poor sample \citepalias{Cooper}. The two pLLSs are indicated by purple asterisks. Histograms for \logu\ and \meh\ are also shown. The darker colors represent only the systems with full ionization solutions, while the limits are shown in the lighter colors. For the histograms we take $\logulim=-3$. The summation of all three tiers is in gray.} 
     \label{fig:mesh}
  \end{center}
\end{figure*}

Figure \ref{fig:mesh} projects our results onto the \logu\ vs. \meh\ plane. All of the LLSs in our sample have heavy-element abundances below $\meh=-1.5$, in sharp contrast to DLAs at similar redshift \citep[][but see \citealp{2015ApJ...800...12C}]{Rafelski2014} and contrary to our expectation of finding a high-metallicity branch analogous to those at $z<1$ \citep{Lehner}.

In Figure \ref{fig:mesh}, best-fit values for \logu~and \meh~are paired with error bars drawn from the MCMC walkers for systems with two-sided bounds on the parameters. For Type 1 upper limits, the systems are shown as lines, and for type 2, as arrows. For each limit, the data allow solutions below and to the left of the line shown, as discussed at the end of Section \ref{subsec.cloudy}. The detections and limits of the LLSs are colored by tier.

The two purple asterisks in the upper left are pLLSs with $\log \nhi < 17.6$. These have the
highest values of \logu, which is not surprising since a system that is more highly ionized will have a small hydrogen neutral fraction and hence a lower \ion{H}{1} column density (also see Figure \ref{fig:mhplot} and Section \ref{sec:nhi}). Interestingly, they also have the lowest bounded values of \meh\ of the entire sample.

Histograms are also shown for \logu\ and \meh\ in Figure \ref{fig:mesh}. The gray histograms represent the entire sample, while the red, blue, and green overlays correspond to the tiered subsamples. Lighter colors indicate limits. In the cases where a limit for the ionization parameter could not be determined, its value was set to the (conservative) minimum value $\logulim=-3$ for the histogram, and the corresponding limit to \meh\ at \logulim\ was then used.

For completeness, we include points from our metal-poor sample \citepalias{Cooper}, shown as black points and limits, observed with the Magellan Echellette Spectrograph \citep[MagE,][]{2008SPIE.7014E.169M} on the 6.5 m Magellan/Clay Telescope. While the metal-poor systems in \citetalias{Cooper} were, by construction, all Tier 1, they overlap completely with the space populated by the Tier 1 and 2 samples from this work. This indicates that LLSs can have low metallicity even in cases where weak metal lines are seen in low-resolution (i.e., SDSS) spectra. However, it may also result from differences in raw data quality between our ESI sample and the sample in \citetalias{Cooper}, as well as differences in analysis techniques (the MagE sample did not float the aluminum relative abundance in the MCMC simulation, which is discussed in Section \ref{sec:arat}).

It is important to note that the Tier 1 vs. 3 separation is less pronounced visually in Figure \ref{fig:mesh} because such a large fraction of the Tier 1 systems have conservative limits and could have much smaller metallicities. Nonetheless, from the histograms of \logu~and \meh~(top and right, Figure \ref{fig:mesh}), there is no obvious trend between tier and \logu\ nor tier with \meh, although the highest bins in both \meh\ and \logu\ are populated only by Tier 3 LLSs (except for one Tier 1 \meh\ upper limit). However, there is a metallicity-ionization space separation between the tiers in the scatter plot. For any given value of \meh, the tiers split from Tier $1 \rightarrow 2 \rightarrow 3$ as \logu\ increases.

This likely reflects our ability to rank LLSs in metallicity by eye using SDSS spectra without any ionization modeling. Since the spectral signature of \ion{C}{4} and \ion{Si}{4} is a doublet with a fairly large oscillator strength, the presence of one of these species in an SDSS spectrum is probably more likely to lead to a ``definitely has metals'' (Tier 3) classification than the presence of weak \ion{C}{2} or \ion{Si}{2}. The triply ionized species' column densities are more dependent on \logu\ than the singly ionized species'. For the test cases of $\logu=-2.2$ and $\logu=-2.6$ discussed in \citetalias{Cooper} (Figure 5), the model column densities for the singly ionized species change on the order of 0.1--0.2 dex, while the  triply ionized species' column densities change by $\sim$0.7--0.9 dex. Thus, our tier classification likely corresponds to regions in metallicity-ionization space, rather than a simple metallicity cut. Using the axes in Figure \ref{fig:mesh}, our categorization of SDSS spectra resulted in a diagonal separation between the tiers, not horizontal as we had anticipated. As described above, the tiers could not be shown to come from separate parent populations in terms of their metallicity.

Our ability to interpret how our results fit into the scheme of galaxy evolution is limited by a lack of context for the LLSs. Without knowing where the absorbers are relative to their host galaxies and what cosmological overdensities they exist within, we cannot straightforwardly determine if they are indicative of the IGM or gas accretion/star formation processes.

\subsubsection{Metallicity and Ionization Trends with \ion{H}{1} Column Density}\label{sec:nhi}

\begin{figure}
  \begin{center}
    \leavevmode
\includegraphics[width=8.9cm]{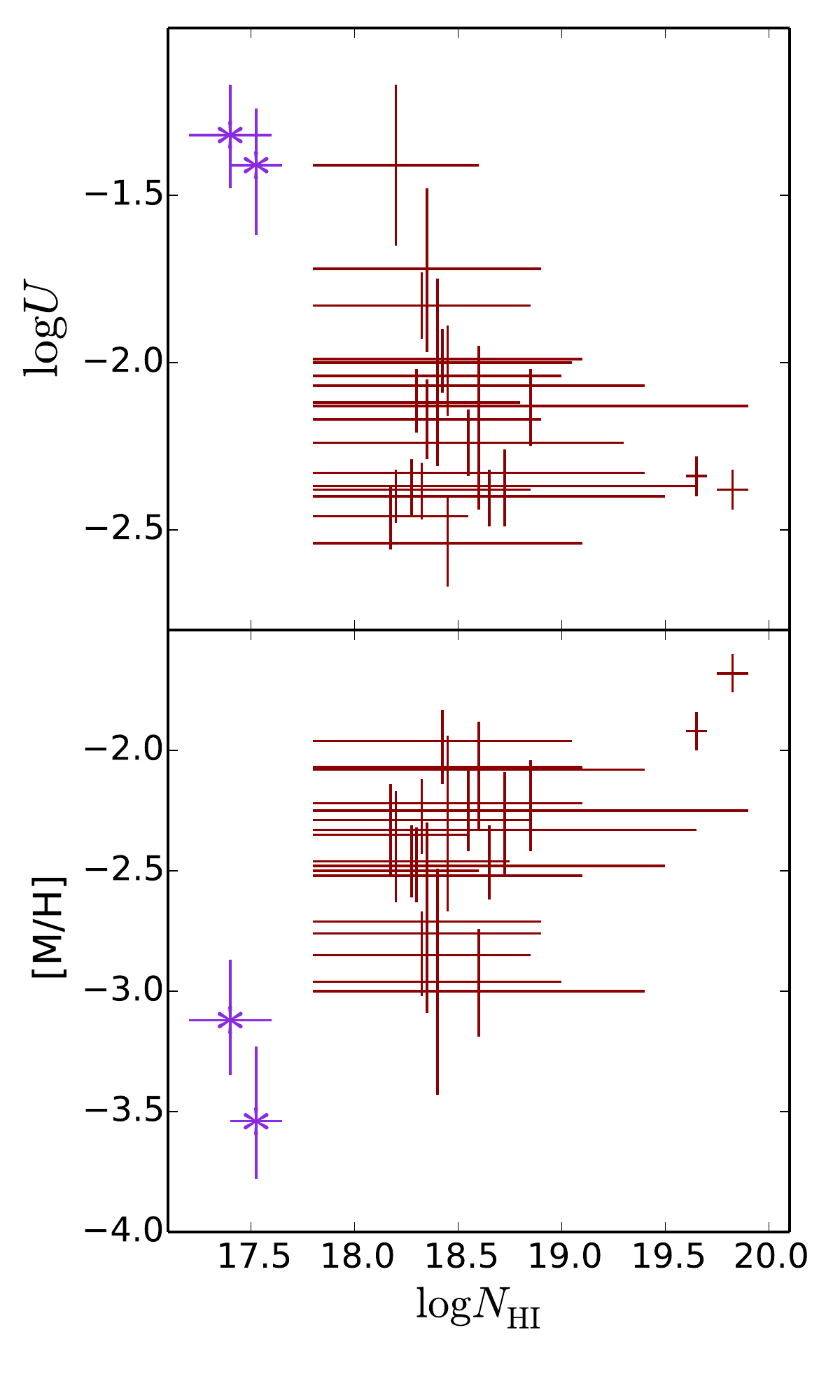}\\
       \caption[mhplot]{Scatter plots comparing \logu\ and \meh\ to $\log\nhi$. For LLSs (shown in red), the plausible range of \nhi\ is indicated by the extent of the horizontal bars. The vertical error bars for \logu\ and \meh\ are placed at the center of the \nhi\ ranges for simplicity, but we emphasize that the crossing point does not indicate a measured value of \nhi. pLLSs, with more accurately measured \nhi, are shown in purple with asterisk markers. The values and error ranges for \logu~and \meh~are from the 4D MCMC, while the range of values for \nhi~are from Voigt profile fits of the Lyman transitions, as described in Section \ref{subsec.cloudy}. Many systems have a minimum \ion{H}{1} column density of $10^{17.8}\cm{-2}$, which corresponds to complete saturation at the Lyman limit at the resolution of our sample. While the data suggest a trend, our sample size is small and \nhi\ is uncertain for most systems.}
     \label{fig:mhplot}
  \end{center}
\end{figure}

Figure \ref{fig:mhplot} shows how \logu~and \meh~compare to \nhi~for the LLSs (red, no markers) and pLLSs (purple asterisks). The error bars for \nhi~show the range of possible values from the Voigt profile fits of the Lyman transitions described above, with the central value of the acceptable \nhi~range chosen for the location of the markers.  Most LLSs are on the flat part of the curve-of-growth and hence have \nhi~uncertain to 1\,dex or more.  We only measure accurate \ion{H}{1} column densities at the low end of our sample (the  pLLSs) and at the high end (which have mild damping wings).

Nonetheless, there is some suggestion of a \nhi-metallicity sequence in these data, although the transition happens in the LLS regime where the tightness of any correlation is masked by uncertainty in \nhi. We find that absorbers with higher neutral fraction (i.e., lower \logu) have larger heavy-element abundances \citep[see also][]{Fumagalli15}, and the pLLSs have the lowest abundances and highest ionization condition. 

The increased abundances found in systems with larger \nhi~suggests a transition to the higher-metallicity DLAs and was also noted in \citet{Fumagalli15}. Firm conclusions on pLLS abundances require a larger sample of pLLSs, as individual examples of pLLSs have also been reported with very high abundances in the immediate vicinity of galaxies \citep{Crighton2013,Crighton}. A more targeted study of pLLSs at these redshifts is also merited by the sample used in \citet{Lehner} to find the metallicity bimodality at $z<1$, since it largely consists of pLLSs. \citeauthor{Lehner} found that the bimodality only exists in systems with $16.2<\log\nhi<18.5$, but found no evidence of dependence on \nhi\ within that group.  \citet{2016arXiv160802584W} verified that at $z<1$ the bimodality does not extend to higher column densities more comparable to those studied in this work.

\subsection{Predicting Metal Column Densities from Model Results}

\begin{figure}
  \begin{center}
    \leavevmode
\includegraphics[width=9.5cm]{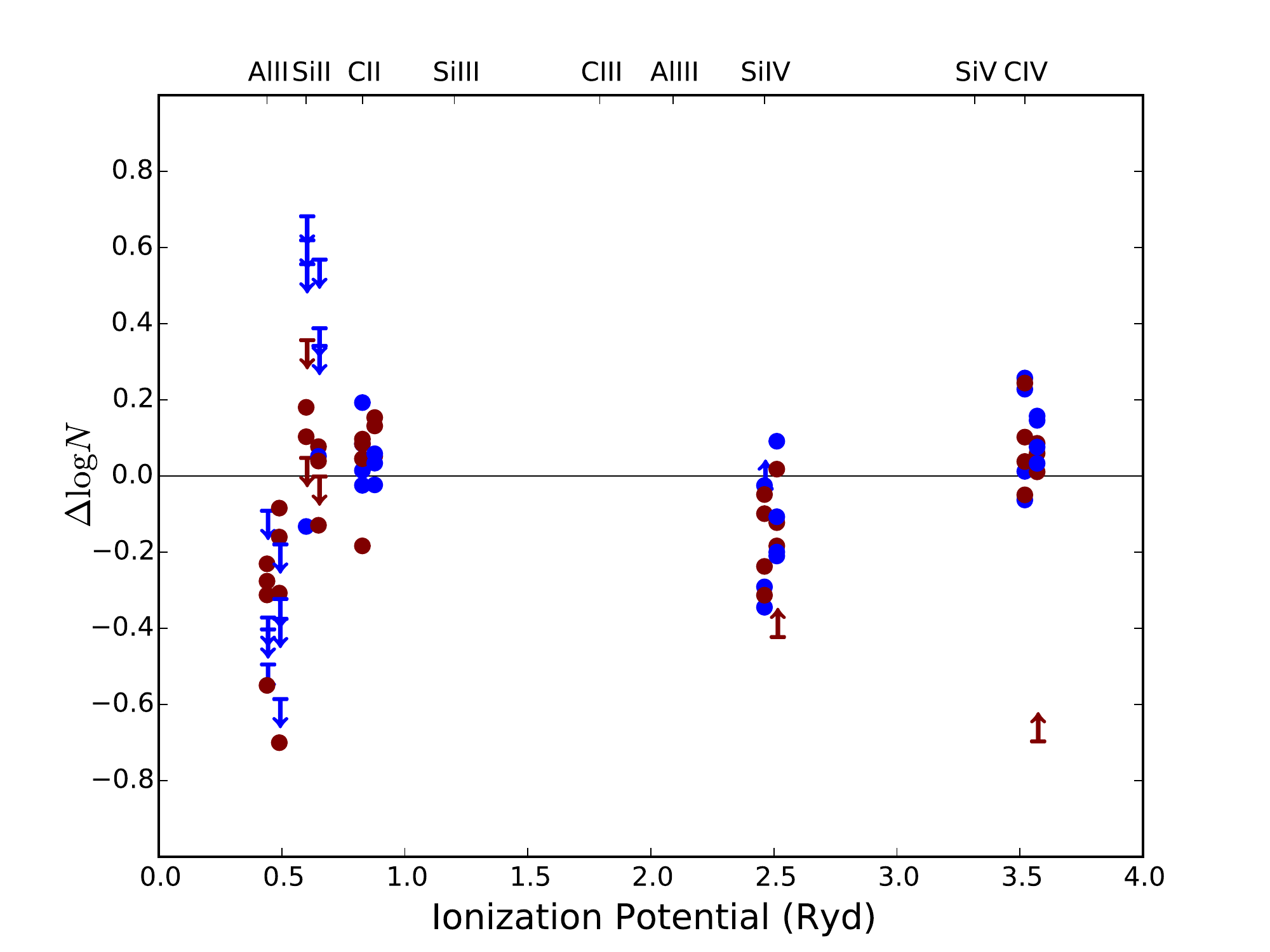}\\
       \caption[Nplot]{Comparison of the measured ionic column densities and those predicted by the median ``best-fit'' ionization models: $\Delta \log N \equiv  \log \N{adpt}-\log \N{pred}$.  Downward facing arrows correspond to a measured column density upper limit. Points corresponding to a system where we measured a \ion{Al}{2} column density are dark red, while those with an \ion{Al}{2} upper limit are blue.  Half of the points are offset slightly along the horizontal axis for graphical clarity. The predicted aluminum values assume a solar-relative abundance pattern. That is, they are taken from a model having $\xh{Al}=\meh$, rather than $\xh{Al}=\meh-\aratio$. Most ions cluster around $\Delta \log N=0$ (horizontal line), with a scatter to either side of $\sim\!0.2\,$dex, while \ion{Al}{2} extends down below $\Delta \log N=-0.5$, with a large number of negative upper limits. The horizontal axis separates the ions by the ionizing potential at which the ion is produced. \ion{Si}{3}, \ion{C}{3}, \ion{Al}{3}, and \ion{Si}{5} have been added for reference. This display scheme is used to show what parts of the ionizing spectrum may be relevant for any discrepancies from zero, such as for \ion{Si}{4} and \ion{C}{4}.}
           \label{fig:Nplot}
  \end{center}
\end{figure}

Some intuition can be gained by examining how well our {\tt Cloudy} models match the measured column densities. We extracted the column densities for each observed ion from the {\tt Cloudy} models (in which the aluminum abundance is effectively a free parameter). The predicted aluminum column densities are what we would expect to measure if aluminum were at solar abundance compared to the $\alpha$-elements. 

These are listed in Table \ref{table:metals}
alongside the actual measurements. This allows us both to confirm that the measured column densities correspond with the model and to predict column densities for species such as \ion{C}{3} and \ion{Si}{3}, since the commonly observed transitions (\ion{C}{3} $\lambda977$ and \ion{Si}{3} $\lambda1206$) could not be reliably measured for our sample (a combination of the dense Ly$\alpha$ forest and the spectral resolution of ESI). This is particularly relevant since \ion{C}{3} and \ion{Si}{3} dominate the carbon and silicon ionization fractions in these LLSs, according to our {\tt Cloudy} grid. 

As shown in Figure \ref{fig:Nplot}, for most ions, the predicted values, \N{pred}, were in fairly good agreement with the measured values, \N{adpt}, though there are some systematic deviations from $\Delta \log N \equiv \log \N{adpt}-\log \N{pred}=0$. While \ion{Si}{4} tends to be under-produced and \ion{C}{4} is slightly over-produced, both are clearly modeled better than \ion{Al}{2}. These discrepancies may be attributable to deficiencies in the ionizing spectrum at the higher energies where these ions exist. Singly and triply ionized species may also be separated spatially within the CGM\footnote{In \citetalias{Cooper}, we show that derived metallicities are robust against \ion{C}{4} and \ion{Si}{4} column density variations, making single cloud models applicable} \citep{2015ApJ...802...10C}. Another plausible explanation for the \ion{Si}{4}/\ion{C}{4} imbalance is non-solar abundance ratios of silicon and carbon. \citet{2016arXiv160802588L} found about half of a sample of LLS and pLLSs from $z\sim$0.1--3.3 have [C/Si] that is non-solar and follows patterns with metallicity similar to those seen in Milky Way stars and DLAs. However, they typically found carbon to be depleted relative to silicon, while we saw the opposite trend: more \ion{C}{4} than \ion{Si}{4} measured then predicted by our solar-abundance pattern ionization models. We suspect our result is more strongly related to the shape of the ionizing spectrum.

The large number of upper limits for \ion{Si}{2} are not alarming, as these correspond to LLSs where the strong \ion{Si}{2} $\lambda 1260$ line is unavailable and we adopt an upper limit from non-detections of weaker lines, usually \ion{Si}{2} $\lambda 1526$. As expected from our prior discussion on the modeling, the \ion{Al}{2} predictions (assuming $\aratio=0$) are discrepant with the observations. Below we quantify the discrepancy of aluminum and discuss possible explanations.

\subsection{Non-solar Aluminum Abundance Ratios}\label{sec:arat}

\begin{figure}
  \begin{center}
    \leavevmode
\includegraphics[width=8.9cm]{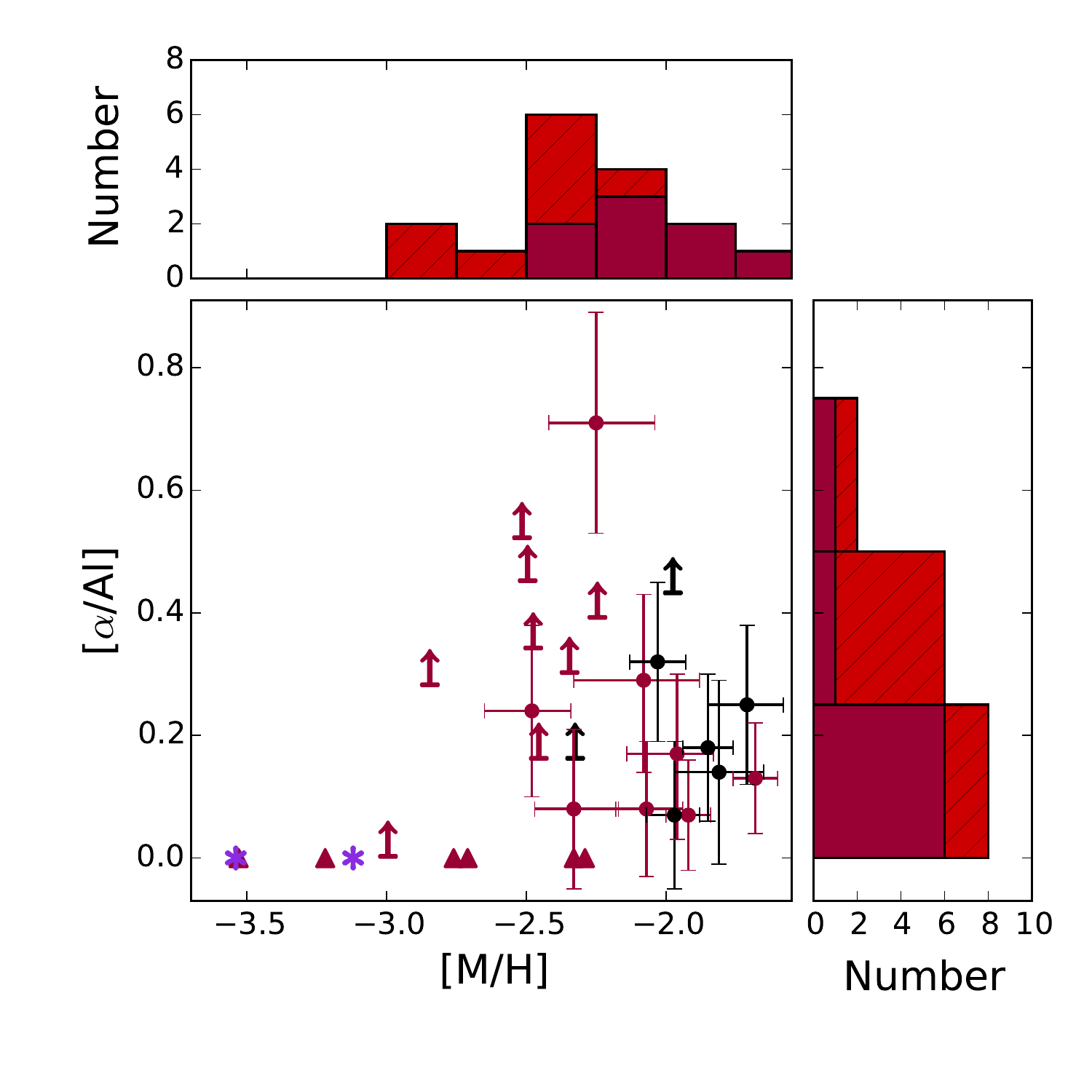}\\
       \caption[alphaplot]{Scatter plot of \aratio\ correction compared to \meh. When models for LLSs included either \ion{Al}{2} and/or \ion{Al}{3}, the \aratio~(points with error bars) or its lower limit (arrows) are shown in red. The systems with measurements cluster with \aratio\ between 0 and 0.4, with one large outlier. Systems with a lower limit to \aratio\ that is below zero are indicated as triangles at $\aratio=0$ (or, for the two pLLSs, asterisks). Black points and arrows are for the LLSs presented in \citetalias{Cooper} (excluding those with negative lower limits to \aratio). The lower limits to \aratio\ tend to be large, with $5/7>0.2$ (excluding those below 0), suggesting that for a large fraction of our sample, aluminum does not have an abundance consistent with a solar-relative abundance pattern.}

     \label{fig:alphaplot}
  \end{center}
\end{figure}

Figure \ref{fig:alphaplot} compares \aratio~from the 4D MCMC for
models that included either \ion{Al}{2} and/or \ion{Al}{3} to
\meh. The arrows indicate lower limits to \aratio, when a value could not be determined
as we only had limits for the aluminum lines, and triangles indicate the fairly uninformative lower limits that are below zero.

For systems where measurements could be made, \aratio\ ranges between 0.07 and 0.71. For most LLSs with detected aluminum ions, we find an \aratio\ that deviates from zero by a small amount, comparable to other ions (see Figure \ref{fig:Nplot}). Of the eight LLSs where we measure an aluminum column density, three systems have $\aratio>0.20$. Additionally, six systems with non-detections have $\aratio>0.3$. For the remaining 21 LLSs, we have either non-detections with small ($<0.2$) or negative lower limits to \aratio\ (12), or no data for \aratio\ (9). The systems with small or negative lower limits are consistent with $\aratio=0$, but could also have large discrepancies, since the actual column densities could be well below the measured limits. Excluding LLSs with a \aratio\ lower limit below zero, we use the Kaplan-Meier estimator (useful for mixed data sets with limits and detections, see Section \ref{sec:pdf}) and find a median aluminum overabundance of $\aratio=0.3$.

We also re-analyzed the LLSs (observed with MagE) from \citetalias{Cooper} using \aratio\ as a model parameter These are shown  in Figure \ref{fig:alphaplot} as black points and arrows. Five of the 17 LLSs selected to be metal-poor in \citetalias{Cooper} are Type 1 upper-limits with \ion{Al}{2} column density upper limits, and \ion{Al}{2} consistent with the metallicity upper-limits derived from other ions. Seven additional LLSs (three from the metal-poor subsample, and four from the metal-blind) have \ion{Al}{2} detections or limits. Three of these have $\aratio<0.2$, and their metallicities changed by $\lesssim0.1$ dex between models with and without \aratio. The two larger measured aluminum detriments are \aratio=0.25 and 0.32. Additionally, two LLS have limits of $\aratio>0.19$ and $\aratio>0.46$. The largest change in metallicity when including \aratio\ as a model parameter is $\sim0.2$ dex, for the LLS with $\aratio>0.46$. The cumulative metallicity distributions of both subsamples presented in \citetalias{Cooper} are minimally affected as only two LLSs have appreciable changes to their metallicities.

Before considering astrophysical interpretations and implications, we note that the accuracy of input atomic physics is an important limiting factor in ionization modeling. Inaccurate atomic ionization and recombination rates can ultimately lead to incorrect column densities for various species. While measuring a total metallicity somewhat marginalizes over this by considering multiple ions, specific abundance ratios are more susceptible to such inaccuracies. In particular, several studies of low-redshift sub-DLAs have found $N_{\rm Al\,III}/N_{\rm Al\,II}$ inconsistent with other measurements  \citep{2001ApJ...557.1007V,2003MNRAS.345..447D,2005A&A...440..819R} and suggest that this may be explained by the dielectronic recombination rate for \ion{Al}{3} to form \ion{Al}{2} being overestimated by as much as 25\% \citep{1986A&AS...64..545N}. \citet{2005A&A...440..819R} also find several components in which aluminum is overabundant relative to carbon by $\sim0.5$ dex; although since they use both \ion{Al}{2} and \ion{Al}{3}, it is unclear how exactly the recombination rate affects this.

To test if this atomic physics uncertainty is the cause of \ion{Al}{2} abundances not matching, we ran a small grid of {\tt Cloudy} simulations in which we decrease the coefficients of the temperature-dependent \ion{Al}{2} dielectronic recombination rate by 25\%. Over a range of \nhi, \meh, and \logu\ representative of our sample, we found that predicted \N{Al II} typically decreases by less than 0.1 dex compared to models run with default atomic physics, with the largest differences being on the order of 0.15 dex. It is clear from Figure \ref{fig:Nplot} that such a change does not alter our result.

In the sample used in this paper, the systems with \aratio\ lower limits (and those with large measured discrepancies) tend toward the lower end of our \meh\ distribution, suggesting that less enriched gas has an aluminum under-abundance. We note that \citet{Crighton2013} measured an \ion{Al}{2} under abundance of 0.3\,dex in a ($\meh=-0.44$) pLLS at $z=2.4$ that is part of a multi-component absorber. \citet{2005A&A...440..819R} found an overabundance of aluminum in a sub-DLA at $z\approx2.2$ that they attributed to incompletely understood dielectric recombination coefficients. \citet{Prochaska2002} found a small enhanced odd-even effect in DLAs with a mean of [Si/Al]$\approx0.4$, but they were unable to correct for dust depletion. We first consider explanations for this signature that do not involve non-solar abundances, then briefly discuss the implications if it is due to elemental abundances.

Refractory elements such as aluminum and silicon are often depleted relative to other elements in systems with large hydrogen neutral fractions (DLAs). This is generally interpreted as due to condensation of these elements onto dust grains \citep[e.g.,][]{Prochaska2002}. While this could explain the aluminum under-abundance, we find no evidence of a similar phenomenon in the silicon abundances and conclude that dust depletion is not a likely explanation. Additionally, \citet{Fumagalli15} find that LLSs typically reside in relatively dust-poor environments.

Another possible resolution to the discrepant aluminum abundances is in modifying the ionizing radiation input to our models. In \citet{Crighton}, the authors performed ionization modeling similar to that used here, with an additional variable that parameterizes the relative contributions of QSOs and galaxies to the radiation field, changing the spectrum of the ionizing radiation. Since \ion{Al}{2} has an ionization potential close to that of \ion{Si}{2}, the model used in \citet{Crighton} predicts that \ion{Al}{2} and \ion{Si}{2} column densities are influenced in the same manner and at roughly the same magnitude by variations in the spectrum of the radiation field, while the column densities of \ion{C}{2} and the triply ionized species are impacted less (by at least an order of magnitude). In our data, \ion{Al}{2} is inconsistent with \ion{Si}{2} in all cases where we find a significant \aratio, and we see no indication that \ion{Si}{2} is inconsistent with any of the other ions except \ion{Al}{2}. Moreover, \citeauthor{Crighton} find that their observations are generally well-fit by small corrections to the nominal spectral shape used in this work \citep[i.e.,][]{haardtmadau}, too small to explain the aluminum discrepancy. Hence, we rule out simple changes to the shape of the ionizing radiation field as the source of the \aratio\ signature.

Assuming that the measured aluminum under-abundance reflects the genuine abundance pattern of the LLS gas, we now consider nucleosynthetic possibilities. Both carbon and silicon are $\alpha$-elements with an even atomic number, while aluminum is odd. A variation in the abundance ratios (relative to solar) of even and odd elements is predicted by some models of hydrostatic burning \citep{Arnett}, and an odd-even effect that is enhanced relative to solar has been noted in the abundance ratios of metal-poor stars \citep{Wheeler}. The odd-even effect can only be measured via aluminum because no other abundant odd-numbered element has an appreciable cross section, except in DLAs. Observations of \ion{Mg}{2} and strong \ion{Fe}{2} can be used to further evaluate chemical abundance ratios and check for dust depletion. While [Si/Al] is expected to display such an enhanced odd-even effect, any claims based on our current data would be premature; [Mg/Al] is another ratio predicted to reflect the signature of an enhanced odd-even effect \citep{Heger}. Complementary to an enhanced odd-even effect, one would expect to see $\alpha$-element enhancement relative to iron typical in gas enriched by Type II supernovae \citep{Wheeler,2014A&A...572A.102F}. Furthermore, while \ion{Mg}{2} and \ion{Fe}{2} have similar dust depletion factors in DLAs \citep{2016arXiv160808621D}, iron is more refractory and its abundance ratios can affirm that depletion is not significant in LLSs. Hence, we favor the interpretation that the aluminum abundance ratio suggests that some metal-poor LLSs represent gas mostly enriched by Type II supernovae. 

\section{LLS Metallicity Distribution}\label{sec:pdf}

The largest motivation for the current work and \citetalias{Cooper} was to assess the metallicity distribution of the LLS population in light of theories of cold-mode accretion. Here, we discuss the
likelihood that LLSs have a metal-poor ``cold-flow'' sub-population tracing inflowing gas \textit{and} an enriched sub-population representing outflowing material.

\subsection{Constructing a Cumulative Distribution from Data Containing Upper Limits}

As our data set consists of a mixture of measured values and limits, we further analyze our distribution of \meh~and \logu~using a form of survival analysis technique to estimate the distribution function. We apply the Kaplan-Meier estimator (KME) for univariate data implemented in \texttt{ASURV} \citep[Rev.~1.2][]{Feigelson,Isobe,Lavalley}. The KME creates a CDF from a mixed data set of detections and upper limits that increases step-wise for each detection and is flat across limits. A detailed discussion of this method can be found in \cite{Simcoe2004}, and \citetalias{Cooper} discusses the validity of the KME to a comparable data set. 

\begin{figure}
  \begin{center}
    \leavevmode
\includegraphics[width=8.9cm]{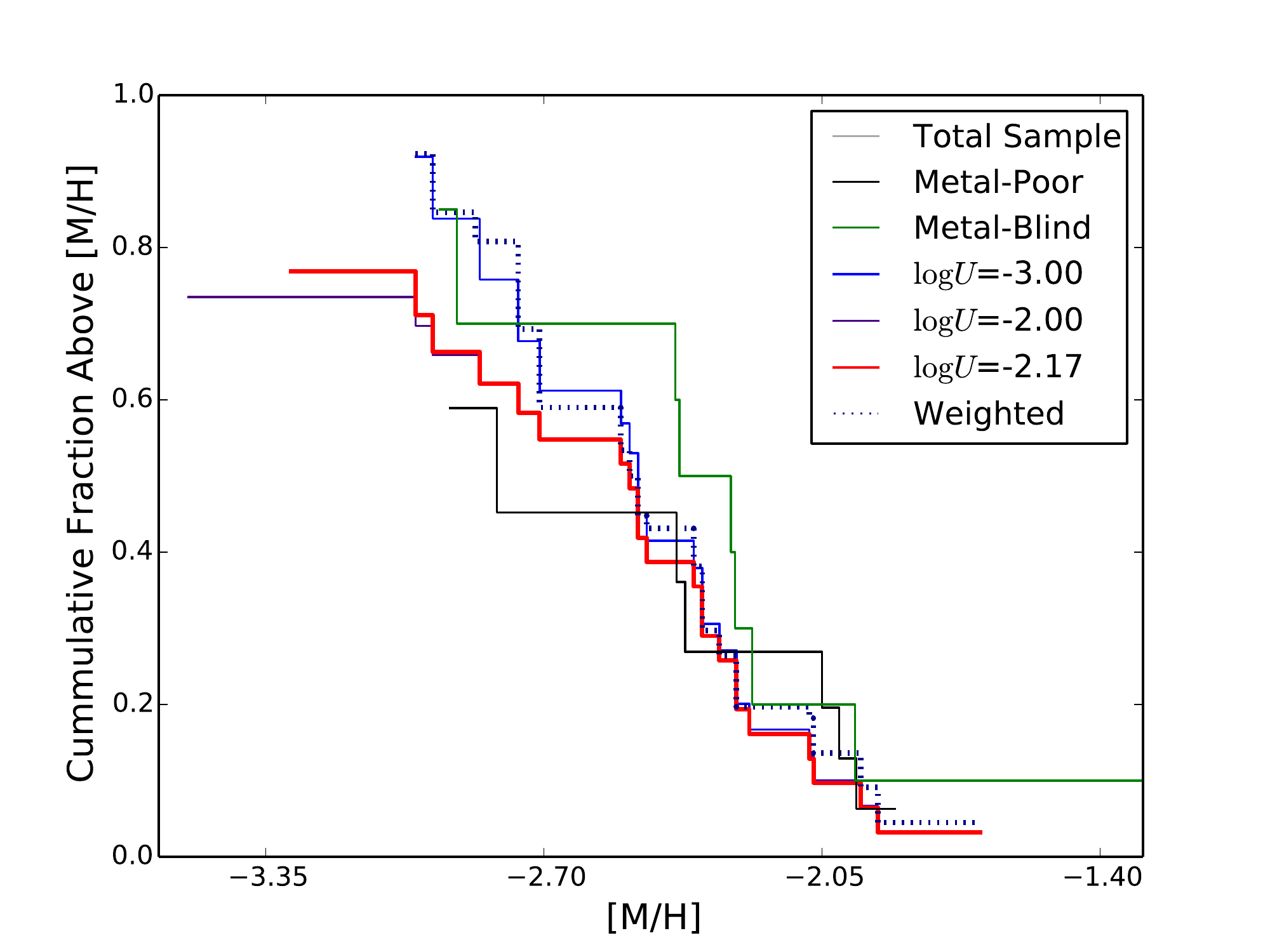}\\
       \caption[cdf]{CDF of LLS metallicities determined using survival statistics. The various colors show how setting \logulim\ changes the CDF. The blue line corresponds to the conservative $\logulim=-3$, while the red and purple are for, respectively, the average of the detections ($\logulim = -2.17$) and a higher value ($\logulim = -2$). The green and black data sets come from the metal-blind and metal-poor samples of \citetalias{Cooper}, respectively. The metal-blind sample has been corrected by $-0.193$\,dex in \meh\ to account for its lower redshift. The dashed blue line shows the CDF with each tier appropriately weighted to correspond to the intrinsic LLS population for $\logulim=-3$.}
     \label{fig:cdf}
  \end{center}
\end{figure}

The step-wise CDF for our data is shown in Figure \ref{fig:cdf}. Three versions of the CDF are shown, corresponding to different choices of \logulim. Since Type 1 \meh\ upper limits become lower with larger values of \logu\ (see Figure \ref{fig:mesh}), increasing \logulim\ for these limits has the effect of causing each upper limit to fall below more detections, which drives the KME to predict a smaller cumulative fraction above the \meh\ value corresponding to each LLS with an ionization solution.

The blue line in Figure \ref{fig:cdf} is for $\logulim=-3$, an overly conservative estimate as seen in Figure \ref{fig:mesh}: none of the detections have \logu~less than $-2.5$, and the values of the limits change appreciably with \logu. The purple line is for $\logulim=-2$, which is an over-estimate, and the red line is for $\logulim=-2.17$, our average measured value of \logu.

It is clear that by removing the step of examining SDSS metals as in \citetalias{Cooper}, we do not recover a missing high-metallicity subsample of LLS (or a bimodal metallicity distribution as seen in \citealt{Lehner}).  In fact, the median metallicity for the sample observed in this paper is somewhat smaller and more statistically significant on account of its larger sample size.  This general result is relatively robust with respect to the choice of \logu\ in our calculation of limits, as evidenced by the large overlap between our CDFs in Figure \ref{fig:cdf}. This is an additional indication that our \meh\ limits reported in Table \ref{table:metals} are extremely conservative upper bounds, since the KME only changes substantially as limits move through the population of detections.

Furthermore, by weighting the tiers, we are able to better recover the intrinsic metallicity distribution function. By including in the KME input each Tier 1 LLS twice and each Tier 3 LLS three times, we obtain a distribution across the tiers that is close to that of the full LLS population. In this input, the distribution is 28\% Tier 1, 13\% Tier 2, and 58\% Tier 3. The full population distribution is 27\% Tier 1, 15\% Tier 2, and 58\% Tier 3, as discussed in Section \ref{sec.obs}. \footnote{This is the full distribution before the DLAs were excluded. However, excluding the DLAs would not have a significant effect on the results.} In Figure \ref{fig:cdf}, this new distribution is shown as the dashed blue line and is also taken at the conservative value of $\logulim=-3$. It is immediately clear that there is very little difference between the weighted and un-weighted distributions, both in blue. This was confirmed using a weighted KME implementation \citep{kme_weight} to account for bias in stratified sampling, which gave a comparable result.

\subsection{Comparison with Other CDFs}

\begin{figure}
\centering
 \includegraphics[width=8.8cm]{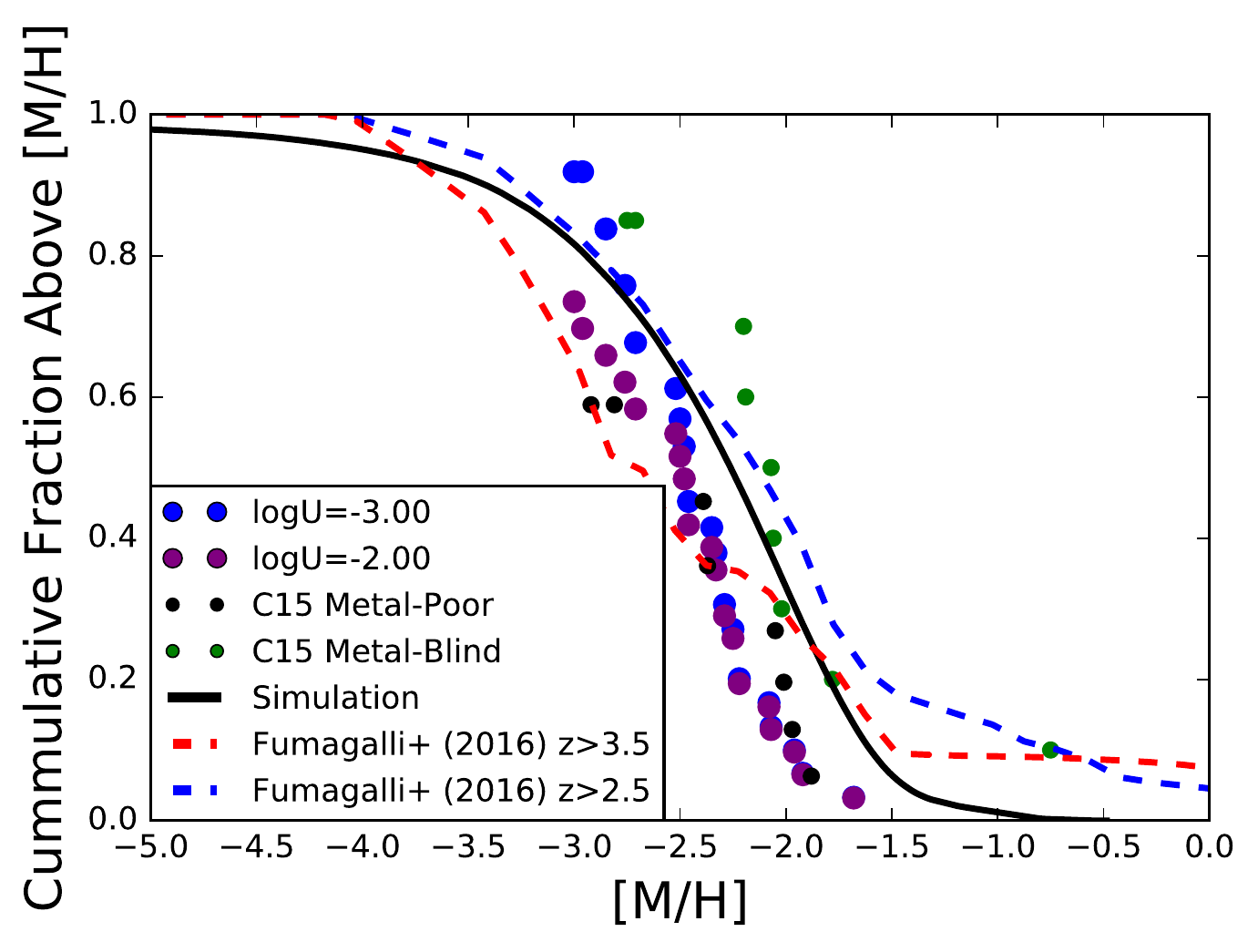}\\
       \caption[simcomp]{Our CDFs derived with $\logulim=-3$ (big blue dots) and $\logulim=-2$ (big purple dots), compared with CDFs from other observational or simulated studies. The smaller black and green dots are the metal-poor and metal-blind CDFs from \citetalias{Cooper}. The solid black curve is the full-volume cosmological hydrodynamic simulation, projected onto a 2D grid \citep{2014MNRAS.445.2313B} and is the same as the one shown in \citetalias{Cooper} for comparison. We also include CDFs of the HD-LLS sample presented in \citet{Fumagalli15} derived from the posterior probability distributions, including only systems having $17.8<\log \nhi<19$ and $z>2.5$ (blue dashed line, 39 LLSs) or $z>3.5$ (red dashed line, 8 LLSs).}
     \label{fig:simcomp}
\end{figure}

In Figure \ref{fig:simcomp}, we compare the CDFs we measured, using $\logulim=-3$ and $\logulim=-2$ with several other observations and a mock CDF constructed from a cosmological simulation. Before discussing this, we note that detailed comparisons are not entirely straightforward as the samples were selected at various redshifts and may have different selection biases, and analysis techniques differ somewhat. Nonetheless, putting the various samples together allows us to coarsely gauge the agreement and variation in high-redshift LLS studies.

The green and black points are the metal-blind and metal-poor samples from \citetalias{Cooper}, respectively, both calculated assuming $\logulim=-3$. As in \citetalias{Cooper}, the metal-blind sample has been shifted by $-0.193\,$dex in metallicity to fairly account for it being at lower redshift ($\overline\zlls=3$) than our other samples, as both the IGM \citep{2011ApJ...738..159S} and DLAs \citep{2012ApJ...755...89R} increase in metallicity with decreasing redshift. The CDF of the metal-poor sample is very similar to that of the broader sample presented in this work, again suggesting that low abundances are commonplace in $z\approx3.7$ LLSs. While the metal-blind sample looks to be enriched by $\sim$0.5 dex relative to other CDFs, it consists of only ten LLSs so sample variance may play an appreciable role.

The CDFs corresponding to the ``high-dispersion'' HD-LLSs presented in \citet{Fumagalli15} are markedly different. Their data set and ours have important contrasting and complementary elements: whereas they use spectra varying in resolution and redshift drawn from previously observed quasars to achieve a large sample size, we selected our objects to control the sample and have increased numbers in a narrow redshift range. We show CDFs for two different cuts to the HD-LLS sample. We do not include the considerable fraction of their sample that has $\nhi>19.0$, but note that they find systematically higher metallicities for such systems. The red dotted curve in Figure \ref{fig:simcomp} is the CDF derived from the eight LLSs in their sample with $z>3.5$ and $17.8<\log\nhi<19$, and the blue dotted curve corresponds to the 39 LLSs with $z>2.5$ in the same \nhi\ range. Notably, both of their CDFs have about 10\% of LLSs with $\meh>-1.5$, much larger than the metallicities found here and in \citetalias{Cooper}  (excluding one LLS in the latter's metal-blind sample). We anticipated finding similarly enriched LLSs in our sample, having removed the metal-poor selection used in \citetalias{Cooper}. 

Looking at metallicities below $\meh=-1.5$, it is not surprising that the HD-LLS $z>2.5$ cut has higher metallicities than the simulation and our observations since about 80\% of the systems included have $z<3.5$. This can be somewhat alleviated by applying the relationship $\meh\propto-0.28z$ used in \citetalias{Cooper} based on IGM and DLA metallicity measurements, and \citet{Fumagalli15} find a slightly larger slope using LLSs with $19.0<\log\nhi<20.3$. We claim that the this sample is in rough agreement with our observations for $\meh\lesssim-1.5$.
 
The HD-LLS $z>3.5$ CDF shows a larger fraction than ours of LLSs at both high and low metallicities, suggesting a broader distribution than we measured. At low metallicities, this may be due to the higher resolution spectra used in their survey allowing for detections of weaker lines or tighter limits and ultimately providing for detections of lower metallicities. For example, in \citetalias{Cooper}, we placed metallicity limits on an LLS of $\meh<-2.7$ using a MagE spectrum and $\meh<-2.9$ using a spectrum with four times higher spectral resolution. While sample variance may also play a role, since the HD-LLS $z>3.5$ cut only includes eight LLSs, it is worth noting that  CDFs based on the HD-LLS sample are intrinsically broader than ours, since they are derived from the full posterior distributions of each LLS, whereas our CDFs are based only on the central values for each LLS. However, as can be seen from Figure \ref{fig:triangle} and Table~\ref{table:ESI}, our posterior \meh\ distributions are not broad enough to account for the high and low metallicity tails seen in the HD-LLS CDF, with 68\% of the posterior probability contained within 0.2--0.3\,dex.  

Despite some small discrepancies between our metallicity distribution and that of \cite{Fumagalli15}, both studies agree on the general distribution, using independent (and differently selected) data sets and analyses. The majority of LLSs (with \meh<19.0) at this redshift have $\meh<-2.0$, without a significant high-metallicity or bimodal population. For comparison, DLAs at $z=3.7$ have a mean metallicity of $\meh=-1.5$ \citep{2012ApJ...755...89R}, and the IGM has median carbon abundance of $\xh{C}=-3.1 (-3.5)$ at $z=2.4 (4.3)$ \citep{2011ApJ...738..159S}.

The solid black curve in Figure \ref{fig:simcomp} is the simulated CDF presented in \citetalias{Cooper}, measured from a full-volume cosmological simulation run using the hydrodynamical simulation code {\tt AREPO}. Further details on the simulation are presented in \citet{2014MNRAS.445.2313B}. Neutral hydrogen and mass-weighted metallicity are projected onto a 2D grid (in slices of 1 Mpc in thickness), and the LLS metallicity distribution is found by treating each projected pixel as an independent line of sight.

The synthetic LLS distribution is over-enriched at the higher end of the metallicity distribution compared to all of our measurements, except for the metal-blind sample in \citetalias{Cooper}. Approaching the lower-end of the \meh\ distribution, the $\logulim=-3$ sample comes into rough agreement with the simulation data at $\meh\lesssim-2.5$. As discussed previously, $\logulim=-3$ for upper limits likely maps to overly conservative \meh\ upper limits, so the LLSs in the simulation are still likely over-enriched relative to what we observe. A negligible fraction of the simulated LLSs have metallicities nearing the largest seen in the HD-LLS sample.

As discussed in \citetalias{Cooper}, two possible explanations for the discrepancy between observed metallicities and the synthetic CDF are: (i) the sightlines probe different parts of the IGM/CGM and\slash or (ii) the winds needed for the simulations to  match observed star formation rates lead to too much enrichment or artificial contamination of relatively pristine material. Investigating the first scenario requires a significant observational program to  identify a related galaxy (or lack thereof) for a large number of LLSs. Regardless of the disagreement in overall enrichment levels, neither the simulated LLSs nor our observations suggest the presence of a metallicity bimodality.
\subsection{Assessing LLS Metallicity Bimodality}

\begin{figure*}
  \begin{center}
    \leavevmode
\includegraphics[width=8.9cm]{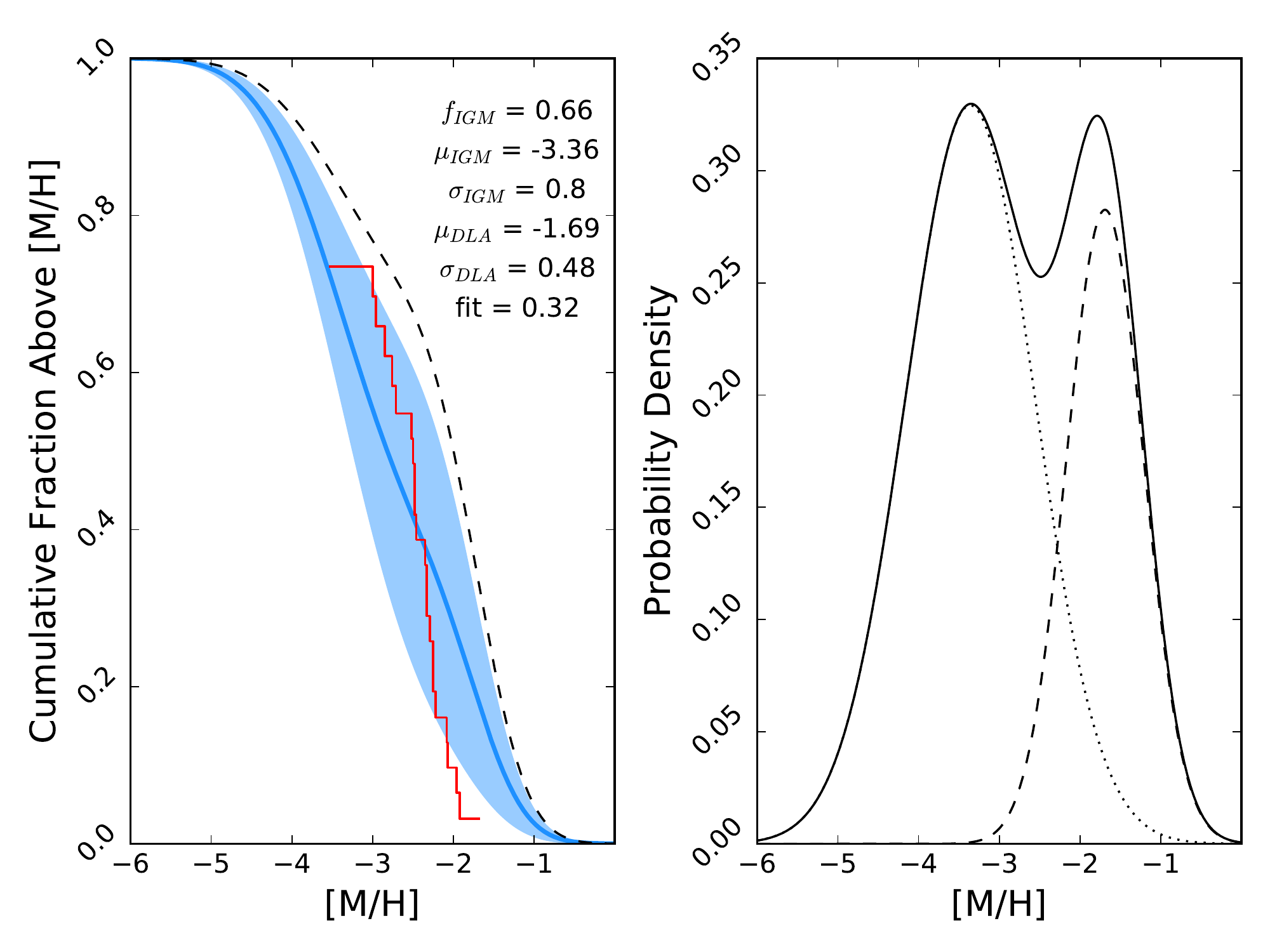}
\includegraphics[width=8.9cm]{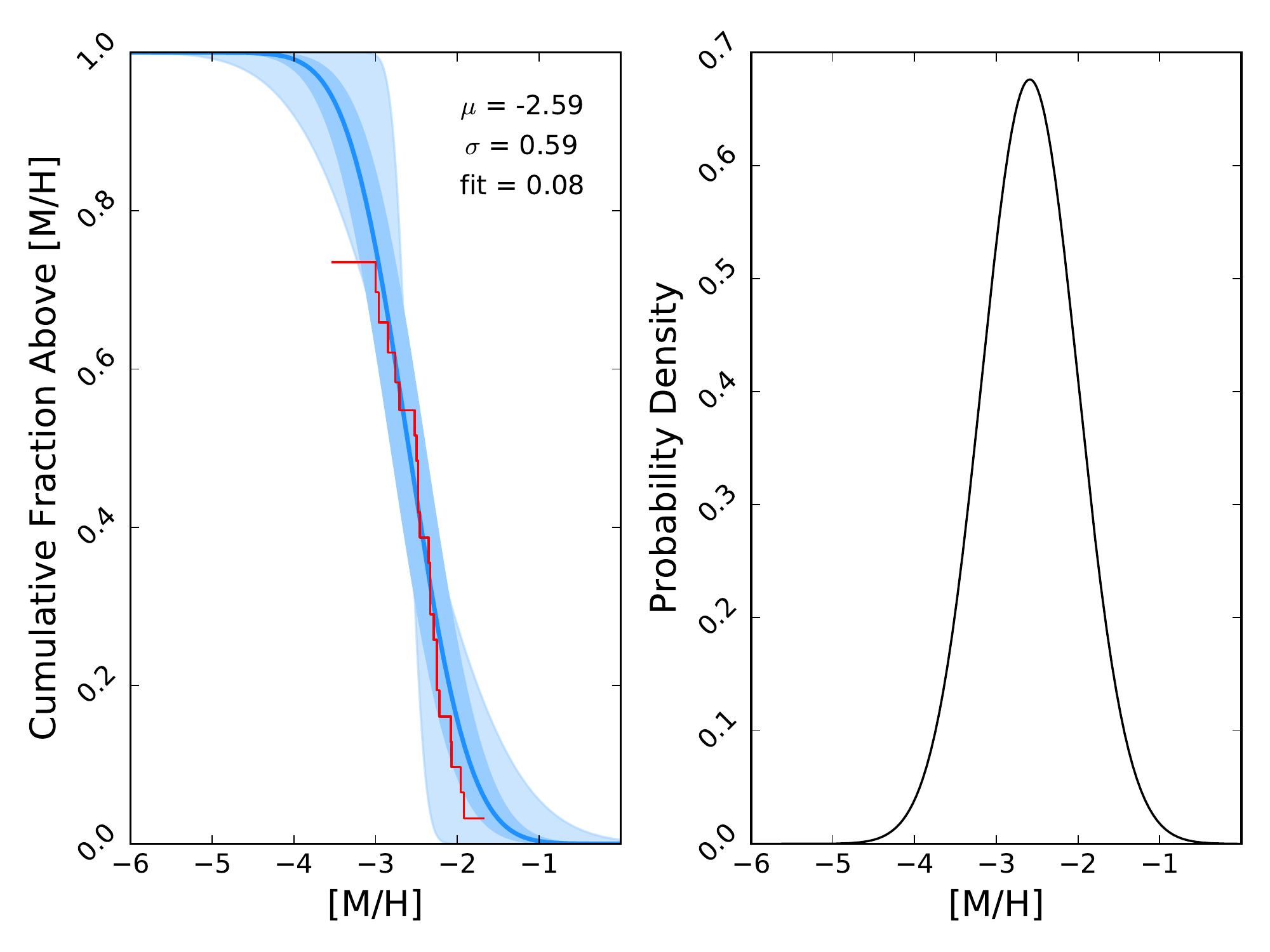}\\
\includegraphics[width=8.9cm]{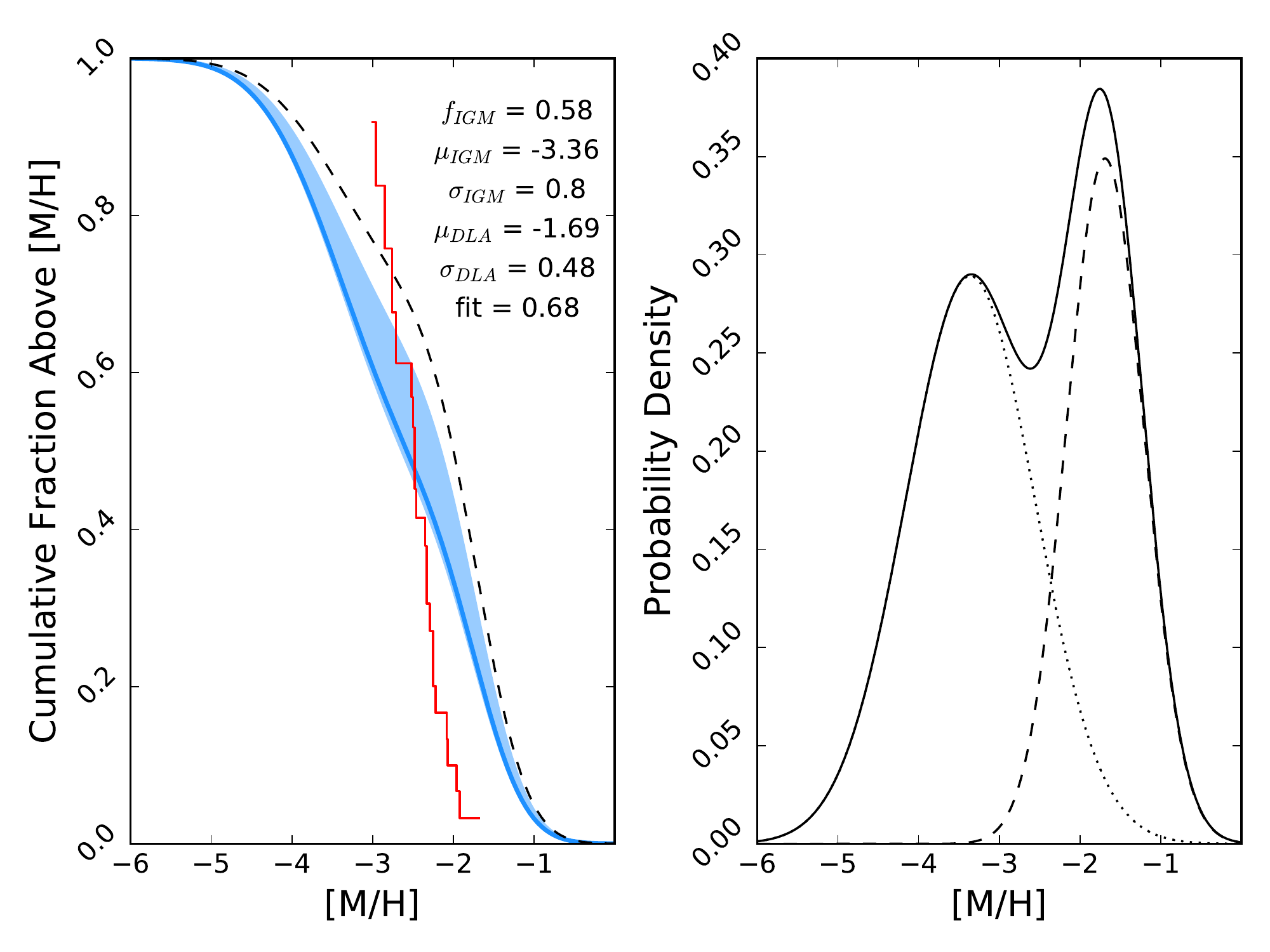}
\includegraphics[width=8.9cm]{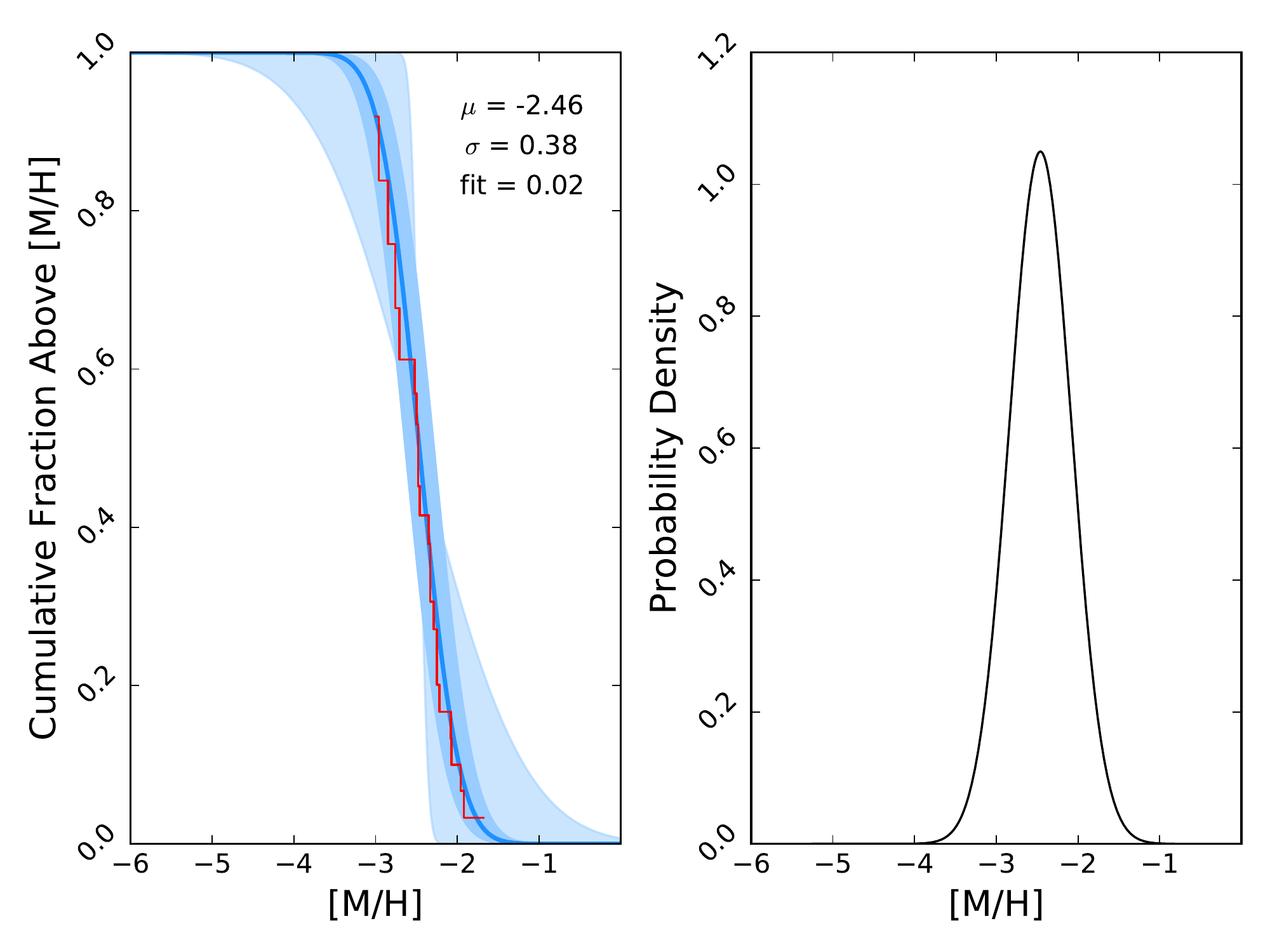}\\
       \caption[mhdist]{Comparisons of the measured CDF with various model CDFs. For each of the four figures, the left panel shows the measured CDF in red  (includes limits) and the model CDF in blue  (1-$\sigma$ error shaded). The right panel shows the corresponding PDF of the best-fit model in solid black with the Gaussian components (if any) as dashed and dotted curves. The black dashed line shows the double-Gaussian model with $f_{\text{IGM}}=0.34$ from \citetalias{Cooper}. The top figures correspond to $\logulim=-2$, while the bottom panels are for $\logulim=-3$. The two left figures show the double-Gaussian model PDF, with metallicities drawn from the $z=3.73$ IGM and DLA with distribution parameters: mean metallicity $\mu_{\rm IGM}=-3.36$ with standard deviation $\sigma_{\rm IGM}=0.8$ and $\mu_{\rm DLA}=-1.69$ and $\sigma_{\rm DLA}=0.48$, respectively. The best-fit fractional contribution by the IGM using linear regression is  $f_{\rm IGM}$, and ``fit'' lists the correspond sum of the squared residuals; lower is better. On the right, we show a PDF of a single Gaussian with a best-fit mean $\mu$ and $\sigma$. Double-Gaussian models do not yield a good fit to the data.}
     \label{fig:mhdist}
  \end{center}
\end{figure*}

We compared the measured CDF derived using survival statistics with various model CDFs, as seen in Figure \ref{fig:mhdist}. As in \citetalias{Cooper}, we fit multi-component models to the data, with strong priors informed by the known abundance distributions of the IGM and DLAs. These represent gas likely to be poorly and highly enriched, respectively, and bracket the LLS \ion{H}{1} column density range.  A key question is whether the bimodal trend seen in LLS abundances at $z < 1$ \citep{Lehner} persists in the early universe at $z \approx 3.5$.

First, we fit a two-component Gaussian model to our data's CDF. The bimodal PDF is described by: $$p(\meh)=f_{\text{IGM}}p_{\text{IGM}}(\meh)+(1-f_{\text{IGM}})p_{\text{DLA}}(\meh),$$ where $f_{\text{IGM}}$ is the fraction of LLSs with metallicities drawn from the IGM metallicity distribution and $(1-f_{\rm IGM})$ is drawn from DLAs. We use DLAs for this fraction since they are thought to consist of material closely associated with galaxies and have metallicities representative of their host galaxies interstellar medium \citep[e.g.,][]{2011ApJ...736...48R}.

 For this model, we use a mean IGM metallicity standard deviation of $\mu_{\rm IGM}= -3.36$ and $\sigma_{\rm IGM}=0.8$, interpolated from measurements at $z=2.4$ and 4.3 in \citet{2011ApJ...738..159S}, and for DLAs we find $\mu_{\rm DLA}=-1.69$ and $\sigma_{\rm DLA}=0.48$ at $z=3.73$ from \citet{2012ApJ...755...89R}. Using a least-squares regression, we found a value of $f_{\text{IGM}}=0.66$ for $\logulim=-2$ and $f_{\text{IGM}}=0.58$ when $\logulim=-3$. As is clearly seen in Figure \ref{fig:mhdist}, this model poorly describes the distribution.  In fact, the majority of our systems have $\meh\approx-2.5$, directly between the mean IGM and DLA metallicities. The observed CDF changes most rapidly in the small valley between the IGM and DLA PDFs, suggesting that the model itself is a flawed representation of the LLS population. Models with the metallicities of the two components as free parameters resulted in either comparably poor fits or one component having negligible contribution to the overall distribution.

In \citetalias{Cooper}, we found that the metallicity distribution of a sample of LLS preselected to be metal-poor could be well-fit by a double-Gaussian model with $f_{\text{IGM}}=0.71$, which corresponds to $f_{\text{IGM}}=0.34$ when extrapolated to the entire LLS population at $z\approx 3.5$. As shown in Figure \ref{fig:mhdist}, this does not accurately model the results for our expanded, more representative sample. In \citetalias{Cooper}, we also modeled a small sample of slightly lower redshift LLSs without any metallicity preselection. Although a bimodal distribution did not provide a high-quality fit, the optimal value of $f_{\text{IGM}}$ agreed within errors bars to the ``metal-poor'' sample. Here, with a larger sample, we establish that a double-Gaussian does not provide a robust fit to the LLS population metallicity distribution at these redshifts, consistent with results showing that the low-redshift bimodality does not extend to larger LLS column densities \citep{2016arXiv160802584W} or to $z=2.3$--3.3 \citep{2016arXiv160802588L}.

In contrast, a single-Gaussian PDF fit very well.
 The best-fit mean metallicity is 
$\mu = -2.29$, with standard deviation $\sigma=0.59$ for
$\logulim=-2$ and $\mu=-2.46$ and $\sigma=0.38$ for
$\logulim=-3$.

Both best-fit mean metallicities lie between the values determined for the IGM and DLAs, but the fit quality is excellent, indicating that  the high-redshift LLS population does not require multiple sub-populations to explain its metallicity distribution.  Rather, LLSs at high redshift appear to be largely metal-poor, although there are examples of LLSs with super-solar abundances \citep{2006ApJ...648L..97P,Fumagalli15}.

However, since our lowest-metallicity systems have upper limits rather than detections \citep[in contrast to][]{Lehner}, there could be a sub-population at extremely low abundance that is missed by the KME. We note that there are several examples in the literature of $z\gtrsim3.5$ LLS with metallicity limits ranging form $\meh\lesssim-3.5$ to $\meh\lesssim-4$ \citep{Fumagalli,Crighton2016}. Generally, such low limits come from high-resolution spectra in which the Ly$\alpha$ forest is resolved well enough to place column density constraints on \ion{C}{3} and \ion{Si}{3}, which dominate the ionization fraction in LLSs at these redshifts (see \N{pred} in Table \ref{table:metals}).  Since such low metallicity limits are difficult to obtain, their fraction of the population remains an open question.

In either case, all the LLSs in our survey have abundances well below those of DLAs and the circumgalactic medium (CGM) of star-forming galaxies at early times. We note that DLAs with abundances similar to those we measure in LLSs do exist, but constitute a small fraction of the DLA population at these redshifts \citep{2015ApJ...800...12C}.

\section{ Summary}

We have analyzed a sample of 33 $z=3.5$--4.2 LLSs to determine their metallicities and abundances. To briefly summarize the methodology, we measure column densities or limits for several ions, then use MCMC techniques and a grid of {\tt Cloudy} ionization models in a 4D (\meh, \logu, \nhi, \aratio) parameter space to determine the heavy-element abundances. Since the neutral hydrogen column density is particularly difficult to measure in LLSs, we opt to marginalize over a range of reasonable values. The main findings of our work are:

\begin{enumerate}
  \item All metallicities are low and range from $-3$ to $-1.68$, resulting in a distribution that is well below that of most DLAs at comparable redshifts \citep{Rafelski2014}. We also find that LLSs are highly ionized, with ionization parameter $\logu>-2.5$ for all systems. Two systems initially classified as LLSs turned out to be pLLSs upon closer inspection and had even lower metallicities ($-3.54$ and $-3.12$) but large $\logu$ ($-1.41$ and $-1.32$, respectively). Coupled with the comparison between LLSs and DLAs, this suggests a moderate trend of increasing metallicity with neutral fraction. However, the literature contains several examples of highly enriched pLLSs in the CGM of a nearby galaxy \citep{Crighton}, so such a relationship may not be straightforward.

\item One-third of the measured aluminum abundances are inconsistent with the measured carbon and silicon abundances (assuming solar-relative abundance ratios), and several aluminum upper limits are several dex below the column densities predicted by our models, requiring us to treat aluminum enrichment as a free parameter in our models. Accounting for lower limits, we find a median aluminum over-abundance of \aratio=0.3. In most cases we cannot determine whether or not aluminum is consistent with solar abundance ratios. Our data suggest a possible trend of lower-metallicity systems having a larger aluminum discrepancy. Although dust depletion is typical in DLAs, LLSs likely reside in environments with less dust, suggesting a nucleosynthetic origin for this observation. Notably, metal-poor stars (expected to have formed at high-redshifts) are often found to have non-solar abundance ratios.

\item We find no hint of a bimodality in the \meh\ CDF recovered from the measurements and limits using survival statistics. Rather, our sample is modeled quite well with a single-Gaussian distribution with a mean metallicity $\meh\approx-2.5$. This is in contrast to the striking bimodality found at $z<1$ in relatively low-\nhi\ systems, and the bimodality that was suggested by an initial sample targeting $z\approx 3.5$ LLSs presumed to be metal-poor based on SDSS spectra \citepalias{Cooper}. It may be that such a bimodality exists via extremely low-metallicity LLSs, but separating them from the bulk of LLSs requires high-quality, high-resolution spectra \citep{Fumagalli,Crighton2016}.

\item Surprisingly, our cumulative distribution is quite similar to that found in the metal-poor sample presented in \citetalias{Cooper}. Our CDF is also roughly consistent with a distribution constructed from the HD-LLS survey \citep{Fumagalli15} at lower redshift (after accounting for redshift differences), although our result is slightly narrower and lacks a small population of highly enriched LLSs. The metallicity distribution constructed from simulations at $z=3.5$, using the code {\tt AREPO}, predicts LLSs that are noticeably more enriched than ours.

\end{enumerate}

The generally low metallicities support the notion that some LLSs at these redshifts represent reservoirs of intergalactic gas that may accrete onto galaxies and fuel star formation. However, without any knowledge of nearby galaxies and/or kinematics, these cannot be definitively classified as cold flows. Furthermore, the lack of a metallicity bimodality does not allow for a simple interpretation where a well-defined fraction of LLSs arise from inflowing or intergalactic gas and the rest outflowing or recycling.
\acknowledgments 

The authors thank the referee C Howk for his thoughtful and thorough review, which greatly enhanced the completeness of this work.

AG acknowledges financial contribution from MIT's Undergraduate Research Opportunities Program (UROP), supported by the FEL UROP Fund. RS and TC gratefully acknowledge support from the National Science Foundation, under award AST-1109915, as well as extended physical support from the A.J. Burgasser Chair in Astrophysics. We appreciate the observational support of the following UH Hilo undergraduates: C~Jones, J~Silva, and I~Trainer. Observing runs were supported by a UH Hilo Research Council Seed Grant.

The data presented herein were obtained at the W.M. Keck Observatory, which is operated as a scientific partnership among the California Institute of Technology, the University of California and the National Aeronautics and Space Administration. The Observatory was made possible by the generous financial support of the W.M. Keck Foundation.

The authors wish to recognize and acknowledge the very significant cultural role and reverence that the summit of Mauna Kea has always had within the indigenous Hawaiian community.  We are most fortunate to have the opportunity to conduct observations from this mountain.

{\it Facilities:} \facility{Sloan}, \facility{Keck:II (ESI)}

\bibliography{ESI_LLS}

\begin{figure*}
\begin{center}
\leavevmode
\includegraphics[width=\textwidth]{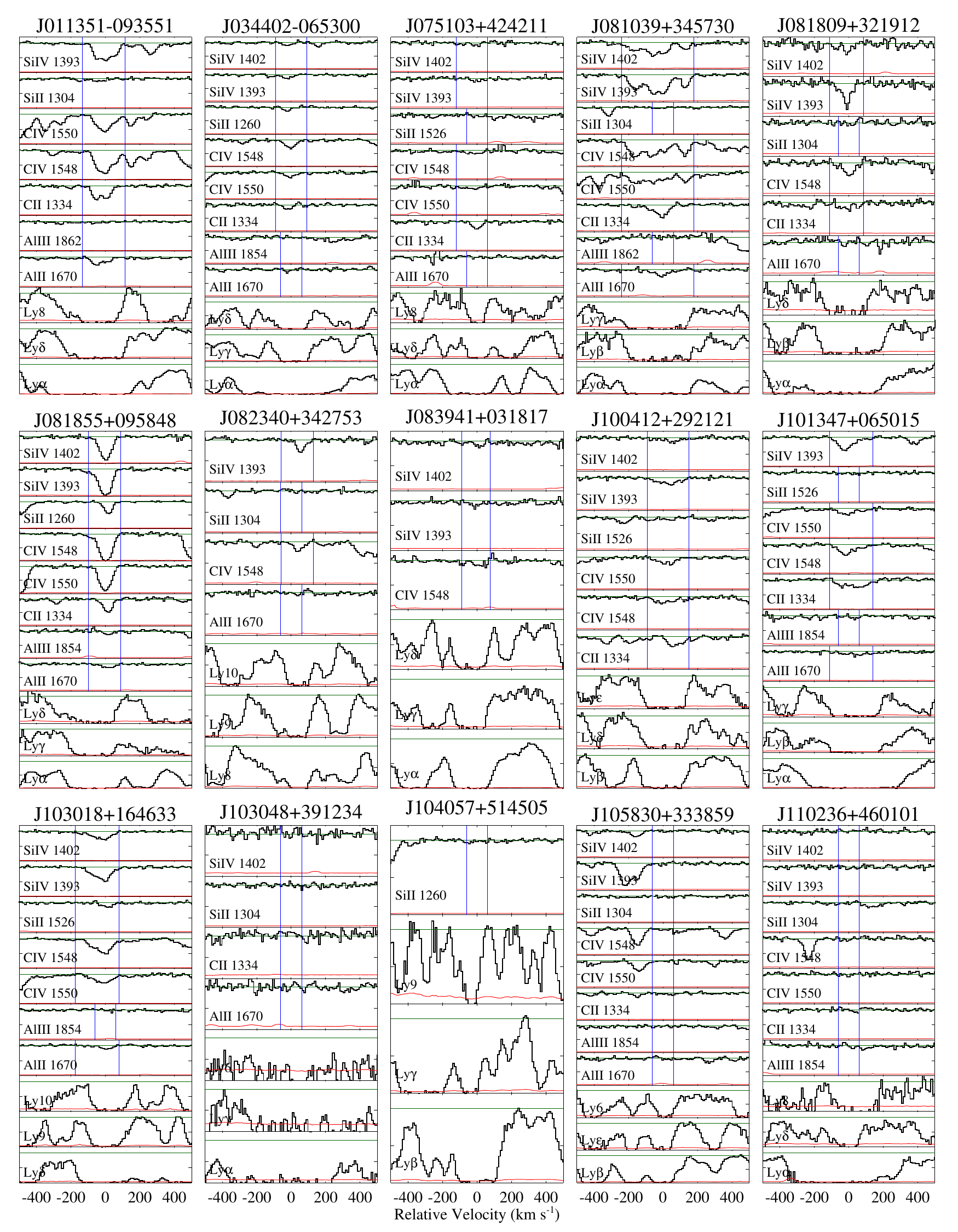}\\
\caption[stack]{The absorption profiles for the 31 LLSs and the 2 pLLSs. For each system, the normalized flux is shown in black at the locations of metal absorption lines that were used for ionization modeling and at a sample of 3 Lyman-series transitions. The 1-$\sigma$ uncertainty on the flux is shown in red. Unity is in green.  The vertical, blue lines show the velocity width over which the column density of the metal lines were measured. In cases where an absorption line could not be detected, the lines show where the 3-$\sigma$ upper limit was determined.}
\label{fig:stack}
\end{center}
\end{figure*}
\addtocounter{figure}{-1}

\begin{figure*}
\begin{center}
\leavevmode
\includegraphics[width=\textwidth]{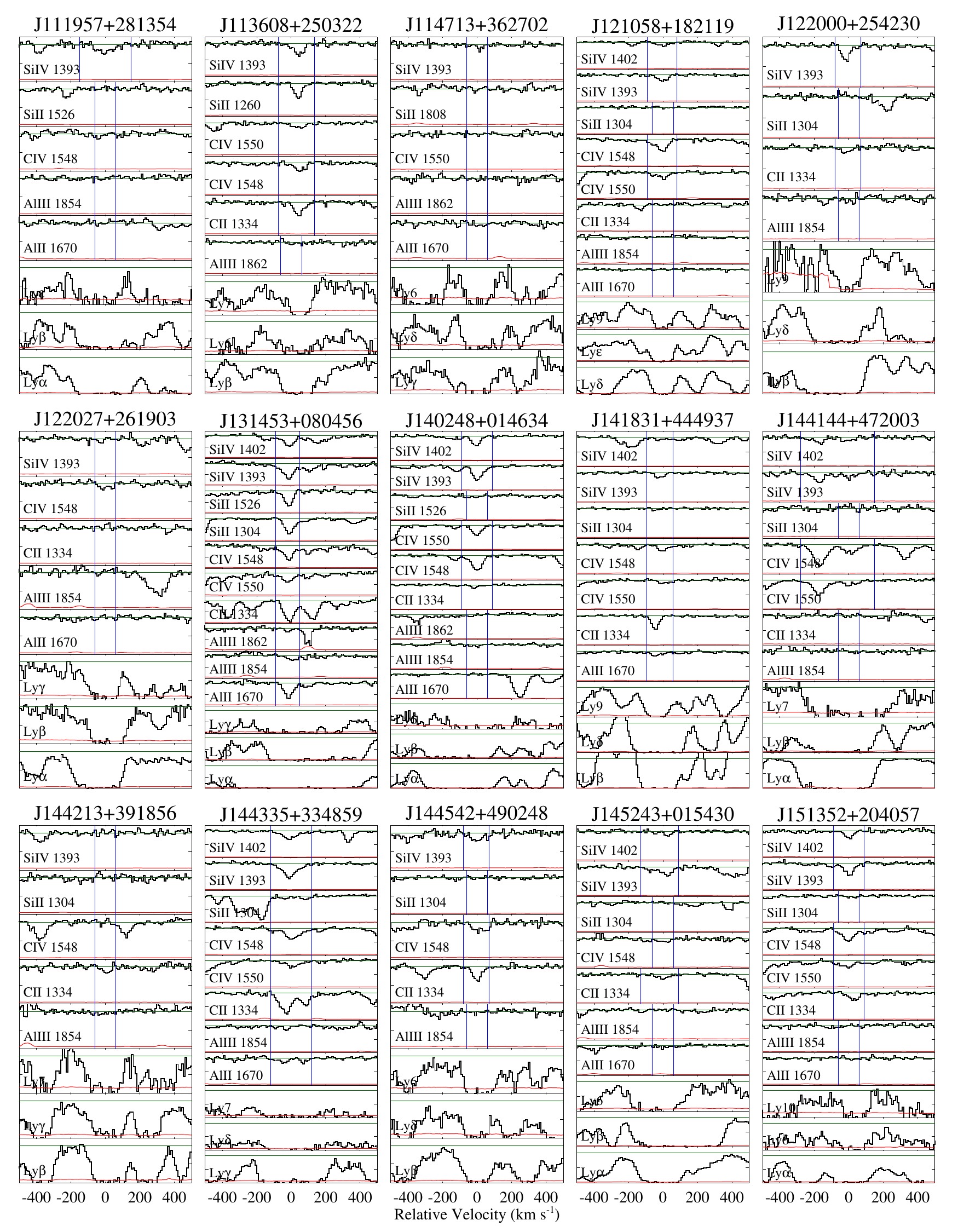}\\
\caption[stack2]{(Continued)}
\end{center}
\end{figure*}
\addtocounter{figure}{-1}

\begin{figure*}
\begin{center}
\leavevmode
\includegraphics[width=0.6\textwidth,clip=false,trim=0cm 15cm 7cm 0cm]{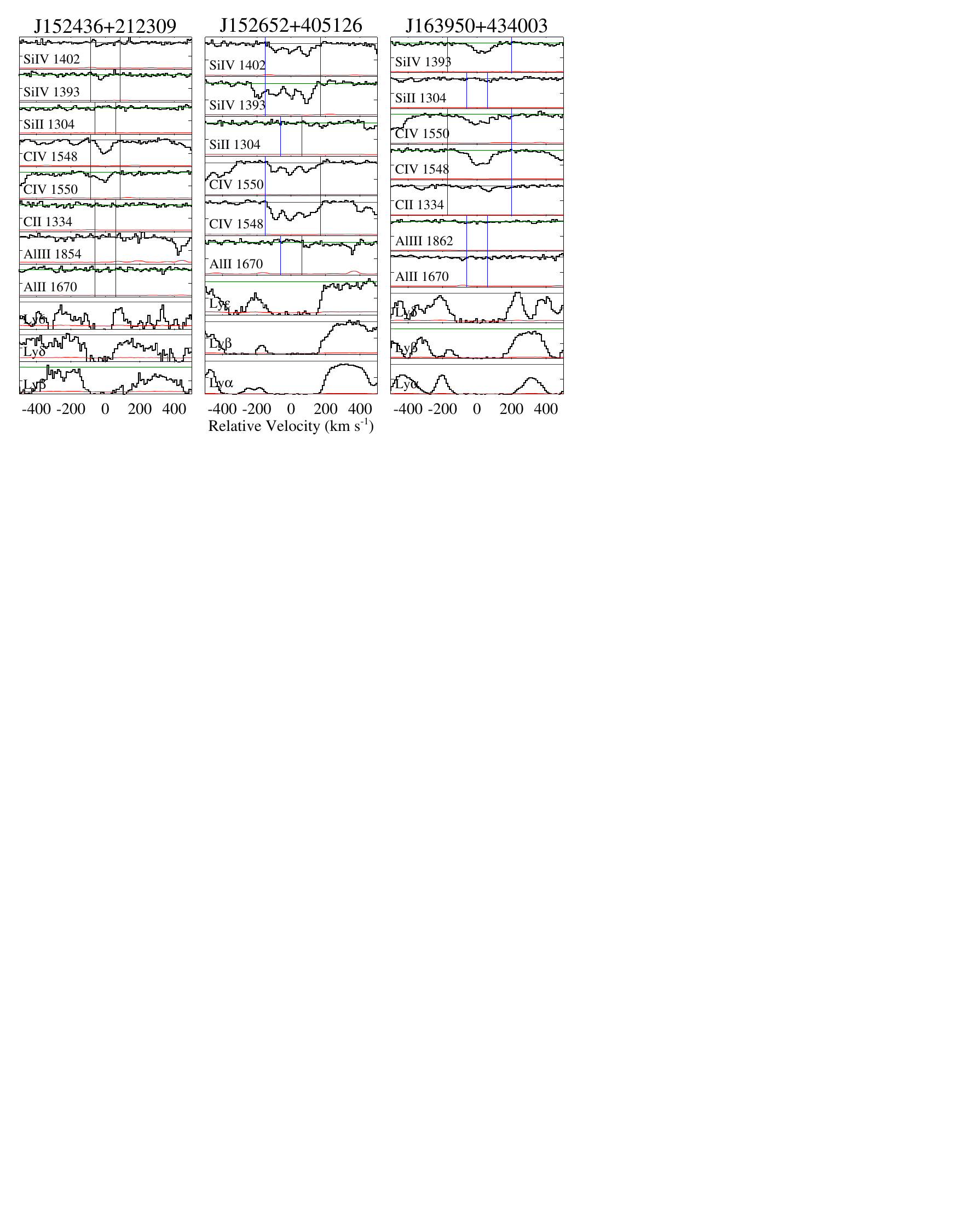}\\
\caption[stack]{(Continued)}
\end{center}
\end{figure*}

\end{document}